\documentclass[a4paper]{revtex4}
\usepackage[T1]{fontenc}
\usepackage[latin1]{inputenc}
\usepackage{graphics}
\draft


\begin{document}

\title{Chaotic Advection near 3-Vortex Collapse }

\author{X. Leoncini}
\affiliation{Courant Institute of Mathematical
Sciences, New York University, 251 Mercer St., New York, NY 10012, USA}
\author{L. Kuznetsov}
\affiliation{Lefschetz Center for Dynamical
Systems, Division of Applied Mathematics, Brown University, Providence, RI 02912,
USA}

\author{G. M. Zaslavsky}

\affiliation{Courant Institute of Mathematical
Sciences, New York University, 251 Mercer St., New York, NY 10012, USA}
\affiliation{Department of Physics, New York
University, 2-4 Washington Place, New York, NY 10003, USA}

\begin{abstract}
Dynamical and statistical properties of tracer advection are studied in a family
of flows produced by three point-vortices of different signs. Tracer dynamics
is analyzed by numerical construction of Poincar\'{e} sections, and is found
to be strongly chaotic: advection pattern in the region around the center of
vorticity is dominated by a well developed stochastic sea, which grows as the
vortex system initial conditions are set closer to those leading to the collapse
of the vortices; at the same time, the islands of regular motion around vortices,
known as vortex cores, shrink. An estimation of the core's radii from the minimum
distance of vortex approach to each other is obtained. Tracer transport was
found to be anomalous: for all of the three numerically investigated cases,
the variance of the tracer distribution grows faster than a linear function
of time, corresponding to a super-diffusive regime. The transport exponent varies
with time decades, implying the presence of multi-fractal transport features.
Yet, its value is never too far from 3/2, indicating some kind of universality.
Statistics of Poincar\'{e} recurrences is non-Poissonian: distributions have
long power-law tails. The anomalous properties of tracer statistics are the
result of the complex structure of the advection phase space, in particular,
of strong stickiness on the boundaries between the regions of chaotic and regular
motion. The role of the different phase space structures involved in this phenomenon
is analyzed. Based on this analysis, a kinetic description is constructed, which
takes into account different time and space scalings by using a fractional equation.

\end{abstract}

\maketitle

\section{Introduction}

The understanding of the advection of passive tracers is of fundamental interest
for many different fields, ranging from a pure mathematical problem, to transport
or mixing related ones. One area extensively studied in the last decade is the
so called chaotic advection\cite{Aref84}-\cite{Crisanti92}. This phenomenon,
resulting from a chaotic nature of Lagrangian trajectories, enhances the mixing
of tracers in laminar flows, while in the absence of chaotic advection the mixing
relies on the much less efficient mechanism of molecular diffusion.

Chaotic advection in geophysical flows is one of the important areas of application,
where the advected quantities vary from the ozone in the stratosphere to various
pollutants in the atmosphere and ocean, or such scalar quantities as temperature
or salinity. The interest in the geophysical flows increases the practical significance
of two-dimensional models and more specifically the advection in the system
of vortices \cite{provenzale99}-\cite{Meleshko93}. In addition to the large
scale geophysical flows, \( 2D \) decaying turbulence is another example, where
the inverse cascade of energy generates coherent structures (vortices) which
dominate the evolution of the flow \cite{Benzi86}-\cite{Carnevale91}. This
type of problems represents only one facet of the interest inherent to the advection
in few-vortices systems. Another facet is related to a transport of advected
particles. It is known from different observations and numerous models, that
the transport of advected particles is anomalous and, in one or another way,
can be linked to the Levy-type processes and their generalizations \cite{Chernikov90}-\cite{Kovalyov2000}.
Although these results pertain to fairly simple flows and models, there are
speculations relating the chaotic advection in low-dimensional flows to particle
dispersion in turbulent flows (see for example a discussion in \cite{zaslavsky93.2}).

The interest to the chaotic advection in a three-vortex system is special not
only for the reasons mentioned above. The three-vortex system is integrable,
and its dynamics can be described in an explicit analytical form. An addition
of a tracer (that can be regarded as another vortex of vanishing circulation)
brings the number of particles in the system to four, which is a minimum number,
from which the point vortex chaos begins \cite{Novikov78,ArefPomph80}; the
relative simplicity of the system permits to study anomalous transport in considerable
detail. A discussion on the importance of three-vortex systems can be found
in \cite{Aref99}.

In this article, we investigate dynamical and statistical properties of the
advection in flows produced by three point vortices with different signs. For
specific conditions on both the initial position of the vortices and their strength,
the collapse of the three vortices to a single point is then possible. 
 We first
summarize the motivations for this work. It is known that for systems involving
a large number of vortices, the local density of vortices fluctuates, these
fluctuations are related to situations when few vortices are close to each other
and form a cluster. The number of vortices included in theses cluster vary,
but the typical number is around 2-3 \cite{Weiss98,Novikov75,Sire99,Zabusky96}
, while cluster involving 4 vortices or
more are much less probable. A given cluster exists only for a definite finite
time \( \Delta t \). During its existence, the notion of space-time structure
of the cluster (or its configuration) can be introduced, and hence the influence
of this structure on typical tracers motion and transport can be studied. Among
different clusters, long time transport properties will be most influenced by
clusters of vortices with the largest lifetime. The typical size of a cluster
can be approximately defined by the minimum distance obtained between two vortices.
The closer the vortices are to each other, the stronger is their mutual interaction
and the longer the cluster lives. In order to change the inter-vortex distances
it is necessary to consider a minimum of three-vortex interactions \cite{Aref79,Aref99,Novikov75}.
And to bring all vortices close together (strong interaction) the configuration between the three
vortices has to be close to to a configuration leading to a collapse of the
vortices \cite{Zabusky96}.  A typical lifetime estimation
of these type of vortices can be assumed to be linked to the period of the motion
of three vortices which are evolving on a close to collapse course. It is shown
in this paper that the closer the vortices are to a collapse configuration the
larger the period of the motion is, and that the growth of the period is exponential
with respect to the closeness to the collapse configuration (see Appendix).
We then expect that some clusters may have arbitrary long lifetime. Having in
mind to shed some light on long time transport properties of systems involving
many vortices, we decided to consider the simpler but nevertheless crucial case
of transport properties in 3-vortex systems for close to collapse motion; this
situations also provides an extreme situation best suited to test a possible
universal transport property for three vortex flows. A simple way to define
how close a motion is from the exact collapse course is through the deviations
of the systems initial parameters from the conditions necessary for collapse.
For instance one condition for collapse is \( K=0 \) (where \( K \) is related
to the angular momentum of the system), a close to collapse motion then necessarily
satisfies \( |K|=\varepsilon \ll 1 \), and \( \varepsilon  \) is the ``distance''
from the collapse condition. Hence a special attention should be made to the
notion of ``close to collapse motion'', as in the whole paper we are referring
as close for a small ``distance'' of the system in parameter space from the
collapse conditions. And we insist that for ``close to collapse'' motions
in the previous sense, no collapse of the vortices occurs and the minimum distance
of approach between the vortices is finite. In this paper the motion of the
vortices is chosen periodic, which retains the possibility of investigating
long time transport properties of these types of flows.
To conduct our study, we use the methodology and the
results of our previous works. The dynamics of tracers is analyzed in a spirit
similar to \cite{KZ98.1}, where the structure of the advection pattern (chaotic
sea, resonant islands, stochastic layers, coherent cores, etc) was investigated
numerically and analytically for the case of three identical vortices. Transport
properties of the advection in that particular case (where no collapse or near-collapse
vortex motion is possible) were found to be anomalous in \cite{KZ2000}. When
vortex circulations have different signs, their dynamics may change considerably;
the collapse phenomenon being one of the most striking examples. The motion
of vortices in the vicinity of the collapse (when the collapse conditions are
just slightly violated) was studied in \cite{LKZ2000}, where different routes
in parameter space all leading to the collapse were outlined. While the kinetics
of advected particles in non-collapsing three-vortex flows was described in
\cite{KZ2000}, the situation with the advection in the near-collapse flows
remained unclear. In this article we describe different topological structures
of the advection pattern, depending on how far is the vortex system from the
collapse conditions. We can speculate, that the characteristics of the transport
of 3-vortex system can be extended to the case of many-vortex flows, since the
advected particle finds itself quite often in the vicinity of a cluster of 3-4
vortices, that define the advection kinetics for a fairly long time-span. For
our study, we chose three different cases, corresponding to a specific route
to collapse: a far from collapse situation, an intermediate one, and one near
the collapse. In these three cases the collapse of the vortices is never reached
and the motion of the vortices is periodic, which allows the study of advection
for large times and a better understanding of transport in three vortex flows,
as we consider here an extreme case with bounded motion, namely the vicinity
of collapse configuration.

In the Section 2, we present the basic equations of the point vortex dynamics,
and discuss the collapse conditions in a three-vortex system. These results
are based on the previous studies of the point vortex collapse \cite{Aref79}-\cite{Kimura90}.
We use the notations introduced in \cite{LKZ2000}, and develop some argumentation
related to our choice of approaching the initial configuration of the vortex
system corresponding to the collapse configuration. Advection equations are
introduced in the Section 3, where we present different tools
used to investigate the dynamic properties of tracers. We focus on Poincar\'{e}
sections, topology of the phase space, and trajectories, while in Section 4
we present different statistical results. They include velocity distributions,
distributions of displacements and its moments, distribution of the Poincar\'{e}
recurrences, etc. An important point of Section 4 is to understand
better how the different regions of phase space influence the statistical properties
of trajectories of advected particles.

On the basis of the obtained statistical information we consider kinetics of
advected particles in Section 5. Fractional kinetics is involved
in that description, and corresponding scalings and characteristic exponents
are estimated on the basis of the results of Section 4. Motivation
for the ``3/2 law'' of the transport is speculated, as well as the multi-fractal
structure of kinetics. Finally, in the Conclusion we discuss different implications
of the obtained results and a possibility to exploit them for the analysis of
the advection in multi-vortex systems.

\section{Near-collapse vortex dynamics}

The evolution of a system of \( N \) point vortices can be described by a Hamiltonian
system of \( N \) interacting particles (see for instance \cite{Lamb45}).
The nature of the interaction depends on the geometry of the domain occupied
by the fluid. For the case of an unbounded plane, the system's evolution writes
\begin{equation}
\label{vortex.eq}
k_{l}\dot{z}_{l}=-2i\frac{\partial H}{\partial \bar{z}_{l}}\: ,\hspace {10mm}\dot{\bar{z}}_{l}=2i\frac{\partial H}{\partial (k_{l}z_{l})}\: ,(l=1,\cdots ,N)
\end{equation}
with the Hamiltonian
\begin{equation}
\label{Hamiltonianvortex}
H=-\frac{1}{2\pi }\sum _{l>m}k_{l}k_{m}\ln |z_{l}-z_{m}|=\frac{1}{4\pi }\ln \Lambda \: ,
\end{equation}
\( z_{l}=x_{l}+iy_{l} \) is the complex coordinate of the vortex \( l \),
\( k_{l} \) its strength and the couple \( (k_{l}z_{l},\bar{z}) \) are the
conjugate variables of the Hamiltonian \( H \), to whom a new energy parameter
is associated 
\begin{equation}
\label{Lambdadef}
\Lambda \equiv e^{4\pi H}=\prod _{l\neq m}|z_{l}-z_{m}|^{k_{l}k_{m}}\: ,
\end{equation}
 in order to simplify subsequent formulae.

The resulting complex velocity field \( v \) is given by the sum of the individual
vortex contributions:
\begin{equation}
\label{velocity_{f}ield}
v(z,t)=\frac{1}{2\pi i}\sum _{l=1}^{N}k_{l}\frac{1}{\bar{z}-\bar{z}_{l}(t)}.
\end{equation}
When \( z_{l} \) evolves according to (\ref{vortex.eq}), \( v \) provides
a solution of the two-dimensional Euler equation, describing the dynamics of
a singular distribution of vorticity 
\begin{equation}
\label{vorticity}
\omega (z)=\sum _{l=1}^{N}k_{l}\delta \left( z-z_{l}(t)\right) .
\end{equation}
 in an ideal incompressible two-dimensional fluid.

The motion equations (\ref{vortex.eq}) have, besides the energy, three other
conserved quantities resulting from the translational and rotational invariance
of \( H \): 
\begin{equation}
\label{constantofmotion1}
Q+iP=\sum ^{N}_{l=1}k_{l}z_{l},\hspace {1.2cm}L^{2}=\sum _{l=1}^{N}k_{l}|z_{l}|^{2}.
\end{equation}
It can be easily verified, that there are three independent first integrals
in involution: \( H \), \( Q^{2}+P^{2} \) and \( L^{2} \), from which it
follows, that the motion of three vortices is always integrable. An analysis
of possible regimes of the motion of three vortices and their classification
can be found in \cite{Aref79,Tavantzis88}.

Among the different types of motion, there is an important special case known
as vortex collapse. This motion is available when the sum of inverse vortex
circulations (harmonic mean) is zero, 
\begin{equation}
\label{strengconditions}
\sum ^{3}_{l=1}\frac{1}{k_{l}}=0
\end{equation}
 and vortex positions positions are such, that the modified angular momentum
\( K \) computed in the reference frame for which the center of vorticity is
placed at the origin, vanishes
\begin{equation}
\label{geomconditions}
K\equiv \left( \sum _{l=1}^{3}k_{l}\right) L^{2}-(Q^{2}+P^{2})=\sum _{l\neq m}^{3}k_{l}k_{m}|z_{i}-z_{j}|^{2}=0
\end{equation}
 Note, that the conditions (\ref{strengconditions}) and (\ref{geomconditions})
do not specify the motion uniquely, rather, they define a range of energies,
for which the collapse is possible (once the values of \( k_{l} \) satisfying
(\ref{strengconditions}) are given). When both of the conditions are satisfied,
the solutions of (\ref{vortex.eq}) are singular and all three vortices collide
at the center of vorticity in a finite time. Depending on the orientation of
the vortex triangle, the collapse time \( t_{c} \) can be either positive or
negative. The first possibility, \( t_{c}>0 \), corresponds to an actual collapse,
while for \( t_{c}<0 \) the vortex configuration expands without bounds. These
two cases are exact images of each other under the time-reversal symmetry, and
we refer to both of them as vortex collapse, keeping in mind, that for \( t_{c}<0 \)
the collapse singularity lies backward in time. During the motion, the vortex
configuration stays similar to the initial one, meanwhile the area of the triangle,
formed by the three vortices grows/decreases linearly in time. We refer the
reader to \cite{Aref79,Tavantzis88,LKZ2000} for more details.

When the collapse conditions (\ref{strengconditions},\ref{geomconditions})
are satisfied only approximately, the motion has a specific ``near-collapse''
type characterized by an emergence of new scales of distances and velocities,
which differ significantly from the initial ones \cite{LKZ2000}. A detailed
analysis of the near-collapse dynamics of three vortices was performed in \cite{LKZ2000}
for the case when two of the vortices have the same strength. A multitude of
motion regimes was found in the vicinity of collapse; different regimes are
distinguished by the way the collapse conditions (\ref{strengconditions},\ref{geomconditions})
are violated, i.e. whether the combinations \( \sum 1/k_{l} \) and \( K \)
are greater, less or equal to zero. Moreover, for some of these combinations,
the motion type also depends on the energy \( \Lambda  \) of the vortex configuration;
for example, when \( \sum 1/k_{l}<0 \) and \( K>0 \) there exist critical
energies \( \Lambda _{c_{1}} \) and \( \Lambda _{c_{2}} \), such that in the
range \( \Lambda _{c_{1}}>\Lambda >\Lambda _{c_{2}} \) the motion has two branches
of periodic motion, for \( \Lambda =\Lambda _{c_{2}} \) the motion is aperiodic,
and for \( \Lambda >\Lambda _{c_{2}} \) there is only one periodic branch left.
A classification of the near-collapse motion regimes, including the behavior
of their length and time scales as the collapse is approached, can be found
in \cite{LKZ2000}.

In order to carry out a detailed study of the advection for close to collapse
situations, we restrict our consideration to a specific one-parameter family
of near-collapsing vortex systems, defined as follows:

\begin{enumerate}
\item Only one of the collapse conditions, the strength condition (\ref{strengconditions}),
is violated; the second condition (\ref{geomconditions}) is satisfied, i.e.
\( K=0 \).
\item The two positive vortices are identical, by an appropriate choice of time units
their strength can be put to 1; in other words, we fix \( k_{1}=k_{2}=1 \).
The circulation of the third vortex is negative, \( k_{3}\equiv -k \), (\( k>0 \)).
In this situation, the collapse happens, when the strength of the third vortex
reaches a critical value \( k=k_{c}\equiv 1/2 \). The ``distance'' of the
system from the collapse can be measured by the amount of the strength detuning
\( \delta  \), 
\begin{equation}
\delta \equiv k_{c}-k=1/2-k
\end{equation}
 We will be considering only the case \( \delta >0 \).
\item The energy of the vortex configuration \( \Lambda  \) is fixed to a constant
value \( \Lambda =0.9 \).
\end{enumerate}
The first two conditions ensure the relative vortex motion to be periodic, and
the choice of energy is such, that in the approach to collapse (\( \delta \rightarrow 0 \)),
the maximum inter-vortex distance grows very slowly (no noticeable changes in
the range of \( \delta  \) considered below) while the minimum inter-vortex
distance rapidly approaches zero. It gives us a convenient ``collapse in a
box'' setting, where vortices initially separated by a distance of order one,
are brought arbitrary close (controlled by \( \delta  \)) together, which may
be of interest for studies of transport in many vortex-systems. For instance,
the ability to bring vortices closer to each other by orders of magnitude makes
these 3-vortex processes a dominant interaction mechanism in a rare gas of vortex
patches \cite{Zabusky96}.

Before proceeding to the discussion of the advection, we will briefly summarize
the results from \cite{LKZ2000}, pertaining to our case. The dynamics of a
three-vortex system with two identical vortices can be mapped to a one-dimensional
Hamiltonian system describing a motion of a particle of a unit mass and zero
total energy in an effective potential, that depends on the strength of the
third vortex \( k \), and the constants of motion \( \Lambda  \) and \( K \)
(see appendix). The dynamical variable \( X \) of this one-dimensional system
is equal to the squared distance between the two positive vortices
\begin{equation}
X\equiv |z_{2}-z_{1}|^{2},
\end{equation}
the other two distances, can be found from the expressions for \( \Lambda  \)
and \( K \). And in our special case (\( K=0) \) the motion is confined between
two single roots of the potential, 
\begin{equation}
X_{min}<X<X_{max}\: ,
\end{equation}
where
\begin{equation}
\label{Xmin}
X_{min}\equiv \left( \frac{1-k}{2k}\right) ^{k/\delta }\Lambda ^{-1/2\delta },\hspace {1.2cm}X_{max}\equiv \left( \frac{1}{2k}\right) ^{k/\delta }\Lambda ^{-1/2\delta }\: ,
\end{equation}
which implies that \( X(t) \) is a periodic function of time, and consequently
the other inter-vortex distances are too, i.e. the relative motion of vortices
in our one-parameter family is always periodic.

As the collapse is approached (\( k\rightarrow 1/2 \), \( \delta \rightarrow 0 \)),
the motion tends to cover all length scales: \( X_{min}\rightarrow 0 \), \( X_{max}\rightarrow \infty  \),
and its period diverges as 
\begin{equation}
\label{periodexponential}
T\sim \frac{1}{\delta }\Lambda ^{-1/2\delta }\: .
\end{equation}
 (see the appendix for the derivation). The influence of this behavior on the
properties of advection is investigated in the next section.

\section{Dynamics of the advection}

A passive particle (tracer) follows the flow according to the advection equation
\begin{equation}
\label{gen.adv}
\dot{z}=v(z,t)
\end{equation}
 where \( z(t) \) represent the tracer trajectory, and \( v \) is the velocity
field. In the case of a point vortex system, the velocity field is given by
Eq. (\ref{velocity_{f}ield}). The incompressibility of the flow allows to write
the advection equation (\ref{gen.adv}) in a Hamiltonian form: 
\begin{equation}
\dot{z}=-i\frac{\partial \Psi }{\partial \bar{z}},\hspace {1.2cm}\dot{\bar{z}}=i\frac{\partial \Psi }{\partial z}
\end{equation}
 where a stream function 
\begin{equation}
\label{stream}
\Psi (z,\bar{z},t)=-\frac{1}{2\pi }\sum _{l=1}^{3}k_{l}\ln |z-z_{l}(t)|
\end{equation}
 acts as a Hamiltonian. This system in non-autonomous, since the stream function
depends on time through the vortex coordinates \( z_{l}(t) \). The character
of this dependence (periodic or not) is important for the further analysis.
Below we will show, that although (\ref{stream}) is quasiperiodic, it can be
made periodic by an appropriate coordinate transformation, which means, that
the advection in our system has a 1 1/2 degrees of freedom Hamiltonian dynamics.

\begin{figure}[!h]
{\par\centering \resizebox*{12cm}{!}{\includegraphics{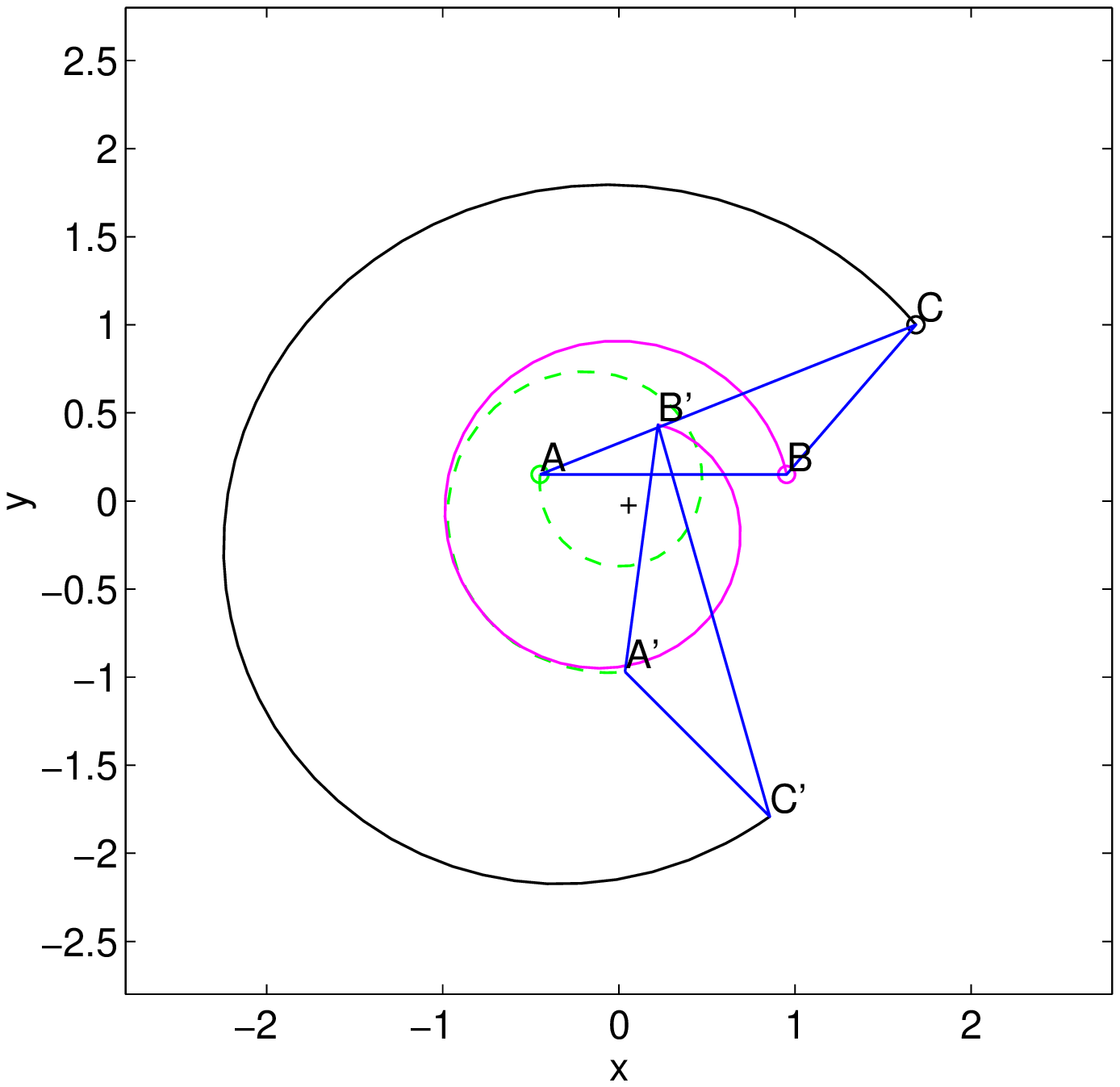}} \par}

\caption{Vortex trajectories for \protect\protect\( k=0.3\protect \protect \) in absolute
reference frame. Initial vortex positions A,B (two positive vortices) and C
(a negative one) are marked with circles. Vortex triangle \protect\protect\( A'B'C'\protect \protect \)
corresponds to \protect\protect\( t=t_{0}+T/2\protect \protect \), it is congruent
to the initial \protect\protect\( \Delta ABC\protect \protect \), but the two
positive vortices are transposed; after another half-period of relative motion
the original orientation restores. Center of vorticity is marked by a ``+''.
\label{v.lab.frame}}
\end{figure}

Indeed, as was mentioned in the previous section, the relative vortex motion
is periodic, i.e. the vortex triangle repeats its shape after a time \( T \).
This does not imply a periodicity of the absolute motion, since the triangle
is rotated by some angle \( \Theta  \) during this time, see Fig. \ref{v.lab.frame}.
In general, \( \Theta  \) is incommensurate with \( 2\pi  \), rendering a
quasiperiodic time dependence of \( z_{l}(t) \).

\begin{figure}[!h]
{\par\centering \resizebox*{12cm}{!}{\includegraphics{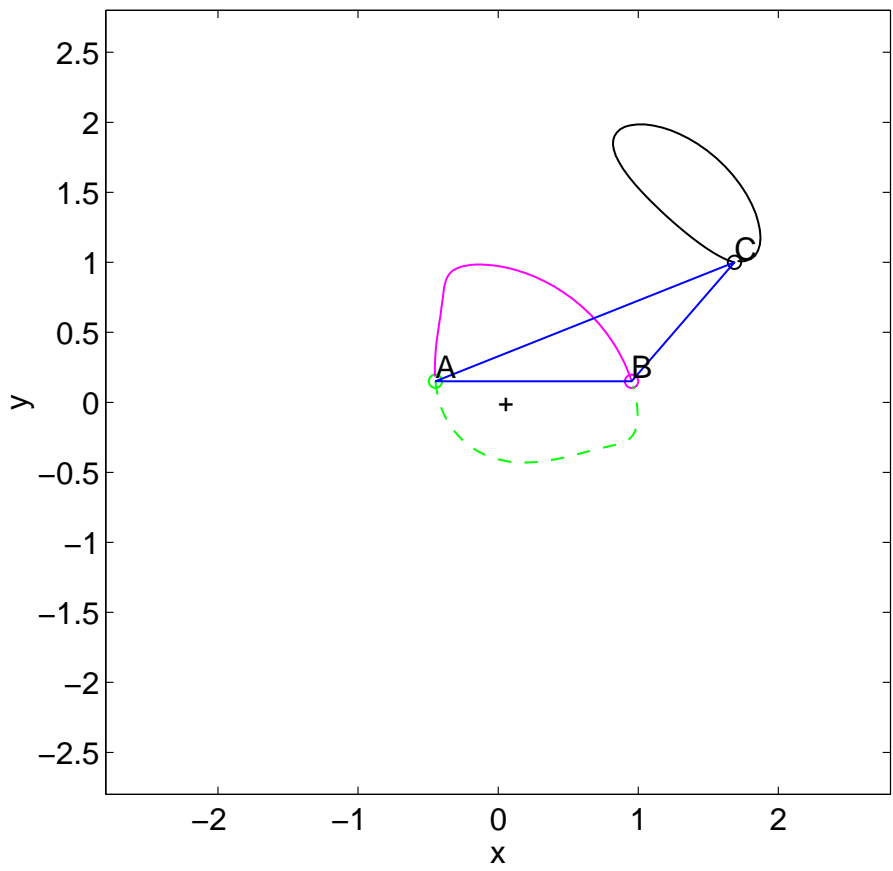}} \par}

\caption{Same trajectories as in Fig. \ref{v.lab.frame}, plotted in the co-rotating
frame: vortex motion is periodic. \label{v.rot.frame}}
\end{figure}

Let us consider a reference frame, rotating around the center of vorticity with
an angular velocity 
\begin{equation}
\Omega \equiv \Theta /T.
\end{equation}
 In this co-rotating reference frame, vortices return to their original positions
in one period of relative motion \( T \), see Fig. \ref{v.rot.frame}, their
new coordinates 
\[
\tilde{z}\equiv z\: e^{-i\Omega t}\: ,\]
 are periodic functions of time.

In the co-rotating frame the advection equation retains its Hamiltonian form
with a new stream function \( \tilde{\Psi } \) which acquires an extra (rotational
energy) term 
\begin{equation}
\tilde{\Psi }\equiv \Psi +\Omega ^{2}/2|z|^{2}.
\end{equation}
 An advantage of this new frame is that \( \tilde{\Psi } \) is time-periodic:
\begin{equation}
\tilde{\Psi }(\tilde{z},\tilde{\bar{z}},t+T)=\tilde{\Psi }(\tilde{z},\tilde{\bar{z}},t)
\end{equation}
 and well-developed techniques for periodically forced Hamiltonian systems can
be used to study its solutions.

Note, that the one-period rotation angle \( \Theta  \) is defined modulo \( 2\pi  \),
making the choice of the co-rotating frame non-unique. We remove this ambiguity
by requiring the negative vortex to make no full revolutions around the center
of vorticity in the co-rotating frame (as in Fig. \ref{v.rot.frame}). This
particular choice of \( \Omega  \) is inconsequential for the further analysis.
The one-period rotation angles, relative motion periods and angular velocities
of the co-rotating frame are presented in the Table \ref{tablesofrotation_freq}.
As the collapse is approached (\( k\rightarrow 1/2 \)), \( T \) grows rapidly
(compare to the formula (\ref{periodexponential})), and the vortices make more
and more turns per period, an acceleration of the rotation speed is also observed.

\begin{table}
{\centering \begin{tabular}{|c|c|c|c|}
\hline 
\( k \)&
\( 0.2 \)&
\( 0.3 \)&
\( 0.41 \)\\
\hline 
\( \Theta (T) \)&
\( -4.18\cdots \approx 2\pi /3-2\pi  \) &
\( -9.4\cdots \approx \pi -4\pi  \)&
\( -28.6\cdots \approx 0.9\pi -10\pi  \)\\
\hline 
\( T \)&
\( 10.71 \)&
\( 17.53 \)&
\( 36.86 \)\\
\hline 
\( \Omega _{v}=\Theta /T \)&
\( -0.39 \)&
\( -0.54 \)&
\( -0.78 \)\\
\hline 
\end{tabular}\par}

\caption{Different values of \protect\( \Theta (T)\protect \) and the associated quantities,
the relative period \protect\( T\protect \) and the resulting apparent rotation
speed.\label{tablesofrotation_freq}}
\end{table}

\begin{figure}[!h]
{\par\centering \resizebox*{12cm}{!}{\includegraphics{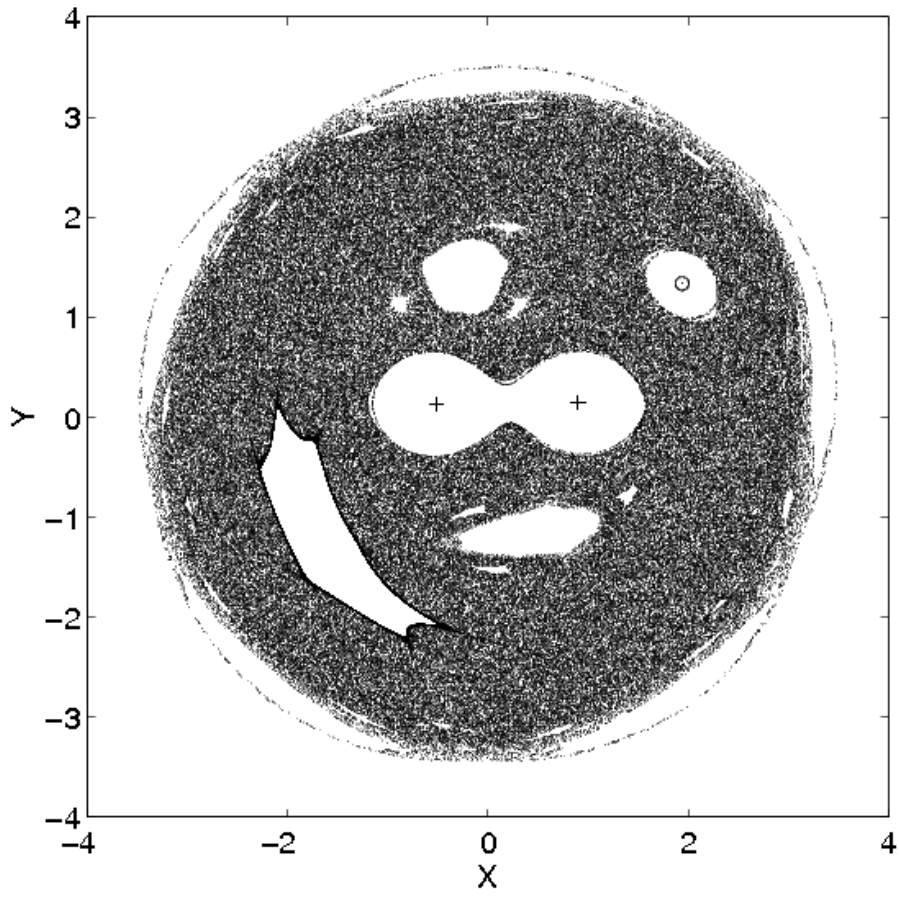}} \par}

\caption{Poincar\'e Map of the system for the far from collapse case. The constants
of motion are \protect\( K=0\protect \), \protect\( \Lambda =0.9\protect \).
Vortex strengths are \protect\( (-0.2,\: 1,\: 1)\protect \). The period of
the motion is \protect\( T=10.73\protect \).\label{poicarresectionk02}}
\end{figure}

\begin{figure}[!h]
{\par\centering \resizebox*{12cm}{!}{\includegraphics{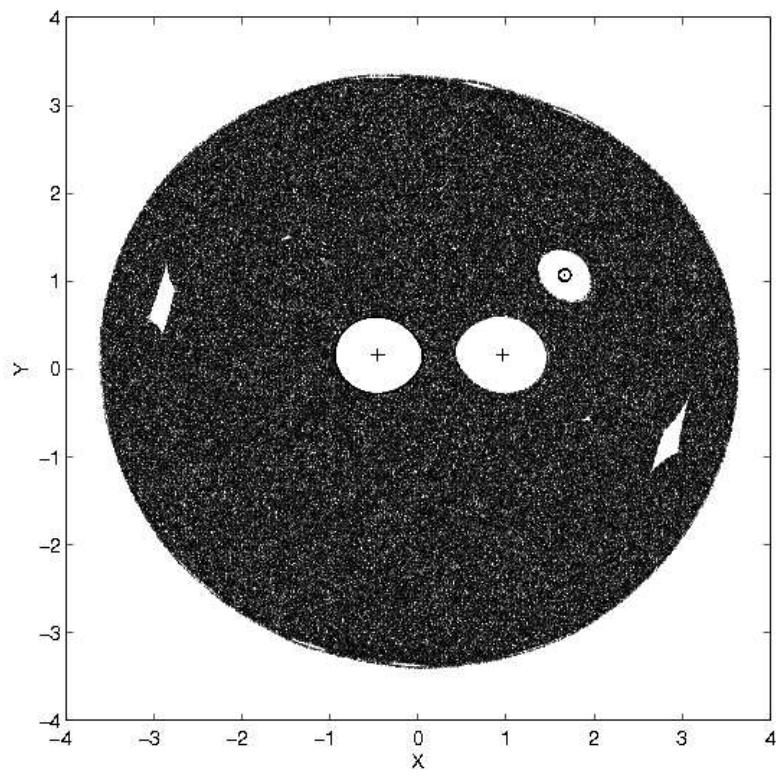}} \par}

\caption{Poincar\'e Map of the system for the intermediate case. The constants of motion
are \protect\( K=0\protect \), \protect\( \Lambda =0.9\protect \). Vortex
strengths are \protect\( (-0.3,\: 1,\: 1)\protect \). The period of the motion
is \protect\( T=17.53\protect \).\label{poinccarek03}}
\end{figure}

\begin{figure}[!h]
{\par\centering \resizebox*{12cm}{!}{\includegraphics{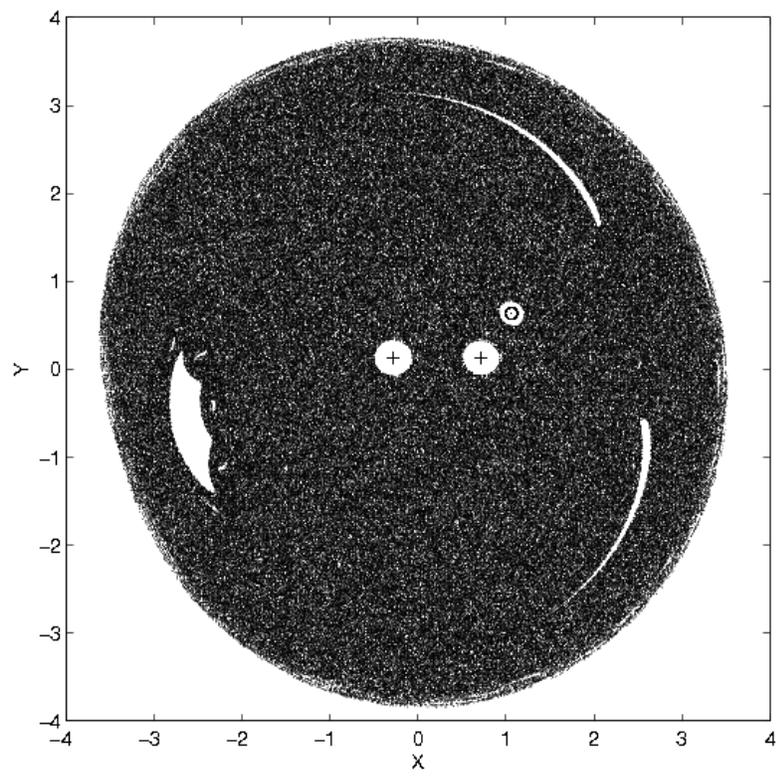}} \par}

\caption{Poincar\'e Map of the system in the close to collapse case. The constants
of motion are \protect\( K=0\protect \), \protect\( \Lambda =0.9\protect \).
Vortex strengths are \protect\( (-0.41,\: 1,\: 1)\protect \). The period of
the motion is \protect\( T=36.86\protect \).\label{poincaresection041}}
\end{figure}

We start our analysis of the advection by numerically constructing Poincar\'{e}
sections of tracer trajectories (in the co-rotating frame). A Poincar\'{e}
section is defined as an orbit of a period-one (Poincar\'{e}) map \( \hat{P} \)
\begin{equation}
\hat{P}z_{0}=\tilde{z}(T,z_{0})=e^{-i\Theta }z(T,z_{0})
\end{equation}
 where \( \tilde{z}(t,z_{0}) \) denotes a solution \( \tilde{z}(t) \) with
an initial condition \( \tilde{z}(t=0)=z_{0} \). Plots of Poincar\'{e} sections
for three different values, \( k=0.2 \), \( k=0.3 \) and \( k=0.41 \) are
shown in Figures~\ref{poicarresectionk02}-\ref{poincaresection041}. Vortex
and tracer trajectories were computed using a symplectic fifth-order Gauss-Legendre
scheme \cite{McLachlan92}. Exact conservation of Poincar\'{e} invariants by
the symplectic scheme suppresses numerical diffusion, yielding high-resolution
phase space portraits.

The Poincar\'{e} sections presented in Figures~\ref{poicarresectionk02}-\ref{poincaresection041}
show an intricate mixture of regions with chaotic and regular tracer dynamics,
typical for periodically forced Hamiltonian systems. All three phase portraits
share common features with the advection patterns, found in a flow due to three
identical point vortices \cite{KZ98.1,NeufeldTel}: the stochastic sea is bounded
by a more or less circular domain, there are a number of islands inside it,
where the tracer's motion is predominantly regular. In particular, all three
vortices are surrounded by robust near-circular islands, known as vortex cores.
Contrary to the case of three identical vortices, where the tracer dynamics
is integrable for a special value of vortex energy (when the vortices form a
steadily rotating isosceles triangle) and has a near-integrable character in
the vicinity, the tracer motion in the near-collapse flow family considered
here is always strongly chaotic, the stochastic sea remains a principal element
of the advection pattern for any \( k \). The phase portraits indicate, that
the degree of chaotization increases with the approach to collapse. For instance,
in the far from collapse case \( k=0.2 \) the islands of regular motion inside
the stochastic sea occupy a considerable area, and as \( k \) increases, their
share drastically diminishes.

A decrease in the radii of the vortex cores is of particular interest, since
they are the robust structures, which also appear in many-vortex systems. The
upper bound on the core radii can be obtained from the minimum approach inter-vortex
distances. The minimum distance between the two positive vortices (see (\ref{Xmin}))
is 
\begin{equation}
R_{1}^{min}=\left( \frac{1-k}{2k}\right) ^{k/2\delta }\Lambda ^{-1/4\delta }\: ,
\end{equation}
at the same moment the distance between the negative vortex and one of the positive
two also reaches its minimum, which is 
\begin{equation}
R_{2}^{min}=\frac{1}{2}\left( \sqrt{\frac{2-k}{k}}-1\right) R_{1}^{min}\: .
\end{equation}
At this moment the vortices are collinear. Since the sum of the core radii of
two vortices cannot exceed the minimum distance between them, we get an upper
bound for the radius of the positive vortex core, \( R^{+}_{core} \) and for
that of the negative one, \( R^{-}_{core} \), in terms of \( R_{1}^{min} \),
\( R_{2}^{min} \): 
\[
R^{+}_{core}=\min \left( \frac{1}{2}R_{1}^{min},\frac{1}{1+\sqrt{k}}R_{2}^{min}\right) \]
\begin{equation}
\label{core.bounds}
R^{-}_{core}=\frac{\sqrt{k}}{1+\sqrt{k}}R_{2}^{min}
\end{equation}
 where we took into account, that the minimum distance between the positive
and the negative vortex is shared between the corresponding cores depending
on relative strength, and choose to define the boundary as the point between
the two vortices with minimum speed. The core radii, measured directly from
the phase portraits in Figures \ref{poicarresectionk02},~\ref{poinccarek03},~\ref{poincaresection041},
and their upper bounds, obtained from (\ref{core.bounds}) are listed in the
Table~\ref{Distances_comparaison}. 
\begin{table}
\centering 
\begin{tabular}{|c|c|c|c|}
\hline 
&
 \( k=0.2 \)&
 \( k=0.3 \)&
 \( k=0.41 \)\\
\hline 
\( r^{+} \)&
 \( \sim 0.51 \)&
 \( \sim 0.43 \)&
 \( \sim 0.16 \)\\
\hline 
\( r^{-} \)&
 \( \sim 0.3 \)&
 \( \sim 0.2 \)&
 \( \sim 0.1 \)\\
\hline 
\( R^{+}_{core} \)&
 \( 0.69 \)&
 \( 0.57 \)&
 \( 0.19 \)\\
\hline 
\( R^{-}_{core} \)&
 \( 0.43 \)&
 \( 0.31 \)&
 \( 0.12 \)\\
\hline 
\( R_{sea} \)&
 \( \sim 3.3 \)&
 \( \sim 3.6 \)&
 \( \sim 4 \)\\
\hline 
\( R_{1}^{max} \)&
 \( 1.48 \)&
 \( 1.68 \)&
 \( 2.11 \) \\
\hline 
\end{tabular}
\par{}

\caption{Comparison of the observed size of the vortex cores \protect\( r^{+}\protect \),
\protect\( r^{-}\protect \) with their estimations \protect\( R^{+}_{core}\protect \),\protect\( R^{-}_{core}\protect \),
given by \ref{core.bounds}; the external ``radius'' of the stochastic sea
\protect\( R_{sea}\protect \), and the maximum distance reached between identical
vortices \protect\( R_{1}^{max}\protect \). \label{Distances_comparaison}}
\end{table}

The presence of islands where the motion is regular, in the stochastic sea,
is known to alter the transport properties of a physical system. This phenomenon
is known as ``stickiness''; when a passive particle, traveling in the stochastic
sea, gets close to an island, it is likely to stick to this island for a while
and mimic a regular trajectory of a trapped particle. And since with each island,
a whole hierarchy of smaller islands around islands is present, the particle
can stick for arbitrary long times, which affects on the whole the transport
properties of the system.

\begin{figure}[!h]
{\par\centering \resizebox*{12cm}{!}{\includegraphics{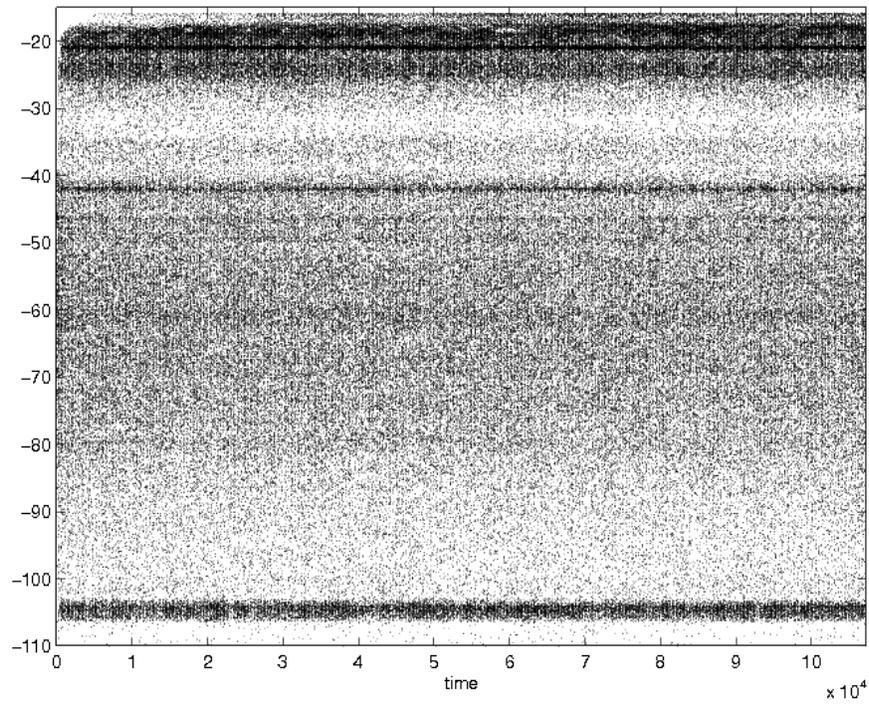}} \par}

\caption{Averaged speed \protect\( V\protect \) over \protect\( 10\protect \) periods
versus time. We notice that some velocities are favored. The distribution does
not seem to be time dependent except at the very beginning. The constants of
motion are \protect\( K=0\protect \), \protect\( \Lambda =0.9\protect \).
Vortex strengths are \protect\( (-0.2,\: 1,\: 1)\protect \). The period of
the motion is \protect\( T=10.73\protect \).
\label{deltathetavstimek02}
}
\end{figure}

In \cite{KZ2000}, stickiness has been exhibited by measuring recurrence times
of particles to a given part of phase space, and plotting the particles position
on the map with a color accordingly to their return times. Particles, that have
long return times, all stick to a particular island, and do not jump from one
to another. Taking these facts into account, we use another way to visualize
stickiness. Indeed, sticking particles have all long coherent time behavior,
which reflects in quantities such as their angular speed, or intrinsic speed.
We decide then to compute the average intrinsic speed over a definite amount
of time of an ensemble of particles and record it. The measured quantity is
the following 
\begin{equation}
\label{average_{s}peed}
V_{i}(m,n)=\frac{1}{nT}\int ^{t_{0}+n(m+1)T}_{t_{0}+nmT}v_{i}(t)dt\: ,
\end{equation}
 where \( n \) is the number of periods over which the speed is averaged, \( m \)
keeps track of the elapsed time, and \( v_{i}(t) \) is the instantaneous speed
at time \( t \) of the particle \( i \). We then define the distribution of
such averaged velocities as 
\begin{equation}
\label{densityofvelocity}
\rho (V;n,m)=\frac{1}{N_{p}}\sum _{i}\delta \left( V-V_{i}(m,n)\right) \: ,
\end{equation}
 and smooth it over an interval to obtain a continuous curve. In fact, as can
be observed in Fig.~\ref{deltathetavstimek02}, for which the speed of particles
is plotted versus time, we can notice that after a brief period of time the
distribution seems stationary, meaning that \( \rho (V,n,m) \) is independent
of \( m \) and therefore we can average its value over \( m \), which in practice
allows better statistics. Figure \ref{deltathetavstimek02} is already informative,
as we can notice some darker stripes, which means that some special average
velocities are favored. This fits with the picture of some particles sticking
to some island for a long time (at least \( >10T \) here). To obtain this data
we computed the trajectories over \( 10000 \) periods for a sample of \( 253 \)
particles and recorded every \( n=10 \) periods. We use the stationarity property
and plot the distribution \( \rho  \) versus the speed \( V \) for the three
different cases \( k=0.2 \), \( 0.3 \), \( 0.41 \), these are represented
respectively in the figures Fig.~\ref{speeddistribk02}, Fig.~\ref{probdistribsk03},
and Fig.~\ref{probdistribsk041}. We notice that the dark stripes observed in
Fig.~\ref{deltathetavstimek02}, correspond to peaks in the density probability.
In order to characterize the origin of these peaks, we plotted in Fig.~\ref{specialspeedsk02},
Fig.~\ref{specialspeedk03}, and Fig.~\ref{specialspeedk041}, the position
of the particles in the phase space contributing to the peaks in the distribution
function. As anticipated, each of the observed peak correspond to a specific
region of the phase space around some island. Concerning the influence of collapse,
we notice that the area occupied by the contributing particles decreases as
the critical condition is approached. We may then speculate that the transport
properties of the three different system, which we discuss in the next section,
may differ in a substantive way.

\section{Anomalous statistical properties of tracers}

Deterministic description of the motion of a passive particle in the mixing
region is impossible, since a local instability produces exponential divergence
of trajectories, and after a short time, the position of a tracer would be completely
unpredictable. Even the outcome of an idealized numerical experiment is non-deterministic
in this situation, since a round-off error is creeping slowly but steadily from
the smallest to the observable scale. For this reason, long-time behavior of
tracer trajectories in the mixing region is usually studied within a probabilistic
approach.

In the absence of long-term correlations, a kinetic description, which uses
Fokker-Plank-Kolmogorov equation and leads to Gaussian statistics, \cite{Zaslavsky92}
works fairly well in many cases. Yet, in the present case, the complex topology
of the advection pattern, illustrated by the Poincar\'{e} sections in Figures
\ref{poicarresectionk02}-\ref{poincaresection041}, indicates that one should
anticipate anomalous statistical properties of the tracers in the chaotic sea.
Singular zones around KAM islands usually produce long-time correlations, which
may result in essential changes in the particle kinetics. Although in some cases
these ``memory effects'' can be accounted for by the modification of the diffusion
coefficient in the FPK equation \cite{Chirikov79}\cite{Rechester80}, often
their influence is more profound \cite{Zaslavsky93}\cite{dCN98}\cite{Castiglione99}\cite{KZ2000},
and leads to a super-diffusive behavior with faster than linear growth of the
particle displacement variance: 
\begin{equation}
\label{anom}
\langle (x-\langle x\rangle )^{2}\rangle \sim t^{\mu }
\end{equation}
 where the transport exponent \( \mu  \) exceeds the Gaussian value: \( \mu >1 \).

In this section we analyze the statistical properties of tracers in the chaotic
region for three vortex flow geometries, introduced above: far from collapse
(\( k=0.2 \), Fig. \ref{poicarresectionk02}), intermediate (\( k=0.3 \),
Fig. \ref{poinccarek03}), and close to collapse (\( k=0.41 \), Fig. \ref{poincaresection041}).
A plot of a time series of the arclength versus time \( s_{i}(t)=\int v_{i}(t)dt \)
for a set of typical tracers trajectories (Fig. \ref{deviationfrommeank02})
reveals an intermittent character of tracer motion: random pieces of trajectory
are interrupted by regular flights, some of which are fairly long. To remain
consistent with previous work, we focus our interest on the character of tracer
rotation, and for that matter, we define its azimuthal coordinate in the center
of vorticity reference frame 
\begin{equation}
\theta (t)\equiv {\mbox {Arg}}\, \, z
\end{equation}
 to be a continuous function of time, i.e. \( \theta (t)\in (-\infty ,\infty ) \)
keeps track of the number of revolutions performed by a tracer. \\

Mean advection angle \( \langle \theta (t)\rangle  \) (here \( \langle \, \, \rangle  \)
denotes ensemble average) grows linearly with time: 
\begin{equation}
\label{mean}
\langle \theta (t)\rangle =\omega t,
\end{equation}
 the values of the average rotation frequency \( \omega  \) for the three cases
are listed in Table \ref{statisticstable}.

The growth of the variance, 
\begin{equation}
\sigma ^{2}(t)\equiv \langle (\theta (t)-\langle \theta (t)\rangle )^{2}\rangle 
\end{equation}
 is faster than linear for all three cases: angular tracer diffusion is anomalous.
From log-log plots of \( \sigma ^{2}(t) \) versus time in Fig. \ref{figsecondmomentvstime},
one may see, that in order to describe the growth of the variance with a power
law 
\begin{equation}
\label{variance}
\sigma ^{2}(t)\sim t^{\mu }
\end{equation}
 one has to introduce different transport exponents for different time ranges.
The values of these exponents, obtained by linear fits of corresponding parts
of the graphs of Fig. \ref{figsecondmomentvstime}, are given in Table \ref{statisticstable}.
Below, the first time range (with the exponent \( \mu _{1} \)) will be referred
to as ``short times'', and the second one (\( \mu _{2} \)) ``long times''. 
\begin{table}
{\centering \begin{tabular}{|l|r|r|r|}
\hline 
&
 \( k=0.2 \)&
 \( k=0.3 \)&
 \( k=0.41 \)\\
\hline 
\( \omega  \)&
\( -0.484 \)&
\( -0.457 \)&
\( -0.387 \)\\
\hline 
\( \mu _{1} \)&
 \hspace{1cm} 1.563 &
 \hspace{1cm} 1.479 &
 \hspace{1cm} 1.679 \\
&
 \( T<3\cdot 10^{4} \)&
 \( T<5\cdot 10^{4} \)&
 \( T<10^{5} \)\\
\hline 
\( \mu _{2} \)&
 1.226 &
 1.707 &
 1.589 \\
&
 \( T>10^{5} \)&
 \( T>1.5\cdot 10^{5} \)&
 \( T>5\cdot 10^{5} \) \\
\hline 
\end{tabular}\par}

\caption{Basic values of transport properties. The average rotation speed of the tracers
and the exponent related to the time evolution of the second moment are given.\label{statisticstable}}
\end{table}

Recently \cite{Castiglione99}, two types of anomalous diffusion were distinguished
by the behavior of the moments, other than variance. The case when the evolution
of all of the moments can be described by a single self-similarity exponent
\( \nu  \) according to 
\begin{equation}
\label{weak}
\langle |x-\langle x\rangle |^{q}\rangle \sim t^{q\nu }
\end{equation}
 was called ``weak anomalous diffusion'', whereas the case when \( \nu  \)
in (\ref{weak}) is not constant, i.e. 
\begin{equation}
\label{strong}
\langle |x-\langle x\rangle |^{q}\rangle \sim t^{q\nu (q)}
\end{equation}
 was named ``strong anomalous diffusion''. The importance of this distinction
comes from the fact, that in the weak case the PDF must evolve in a self-similar
way: 
\begin{equation}
\label{selfsim}
P(x,t)=t^{-\nu }f(\xi ),\hspace {1cm}\xi \equiv t^{-\nu }(x-\langle x\rangle )
\end{equation}
 whereas non-constant \( \nu (q) \) in (\ref{strong}) precludes such self-similarity.
Note, that a self-similar PDF evolution can have a more general form than (\ref{selfsim}),
with \( t^{-\nu } \) replaced by an arbitrary decaying function of time \( g(t) \):
\begin{equation}
\label{selfsim2}
P(x,t)=g(t)f(g(t)(x-\langle x\rangle )),
\end{equation}
 which means, that if for different time decades \( g(t) \) has different asymptotics,
the self-similarity exponent \( \nu  \) will change from one decade to another.
This variation of \( \nu  \) with time (in particular, differences of transport
exponents for short and long times in Table \ref{statisticstable}) is not related
to the type (strong or weak) of anomalous diffusion.

We have performed the measurements of a set of moments of tracer angular PDF
(including non-integer values of \( q \)) defined as: 
\begin{equation}
M_{q}(t)\equiv \langle |\theta (t)-\langle \theta (t)\rangle |^{q}\rangle 
\end{equation}
 for the three vortex geometries. Time evolution of each moment was fitted by
a power law: 
\begin{equation}
\label{mgrowth}
M_{q}(t)\sim t^{\mu (q)}
\end{equation}
 separately for short, and for long times. The results are summarized in Figures
\ref{exponentssmalltimesk02}, \ref{Figexpvsmomsmalltk03}, \ref{Figexpvsmomsmaltk041}
(short times) and \ref{exponentslongtimesk02}, \ref{Figexpvsmomlongtk03},
\ref{Figexpvsmomlongtk041} (long times), where the exponents \( \mu (q) \)
are plotted versus the moment number \( q \). In all cases, the apparent absence
of a single linear fit indicates the presence of strong anomalous diffusion.
This property was also found in \cite{KZ2000} (by comparison of the scaling
properties of the central part of tracer PDF with the behavior of the variance)
in a flow due to three vortices of equal strength. Thus, strong anomalous diffusion
is a generic property of advection in three vortex flows.

Our results show, that \( \mu (q) \) is well approximated by a piecewise linear
function of the form:
\begin{equation}
\label{Cform}
\mu (q)=\left\{ \begin{array}{ccc}
\nu q & {\mbox {for}} & q<q_{c}\\
q-c & {\mbox {if}} & q>q_{c}
\end{array}\right. 
\end{equation}
 where \( c \) is a constant, and \( q_{c} \) is a crossover moment number
\( q_{c}=c/(1-\nu ) \). In \cite{Castiglione99}, where this form was introduced,
it was found, that it fits fairly well the numerically obtained values of \( \mu (q) \)
in all cases of strong anomalous diffusion, considered there, although a theoretical
example of a system with arbitrary (concave) \( \mu (q) \) was mentioned. Note,
that deviations from (\ref{Cform}), occurring in the crossover region \( q\approx q_{c} \),
are probably a result of finite observation time, and the form (\ref{Cform})
might be precise in the limit \( t\rightarrow \infty  \).

As we have mentioned, the non-constant \( \nu (q) \) in (\ref{strong}) is
incompatible with the self-similar evolution of tracer distribution. Let us
introduce an ``almost self-similar distribution'' 
\begin{equation}
\label{almost}
P(x,t)=\left\{ \begin{array}{ccc}
t^{-\nu }f(t^{-\nu }x) & {\mbox {for}} & x\ll vt\\
0 & {\mbox {if}} & x>vt
\end{array}\right. 
\end{equation}
where the exact self-similarity (\ref{selfsim}) is broken only in the time-dependence
of the cut-off. This modification of the exact self-similarity relation (\ref{selfsim})
takes into account the fact, that tracer speed is bounded by a maximum speed
\( v \). If \( f(\xi ) \) decays sufficiently fast at infinity (e.g. like
a Gaussian), the cutoff behavior is irrelevant and the moments of (\ref{almost})
follow (\ref{weak}), but if \( f(\xi ) \) has a power tail, the almost self-similar
distribution (\ref{almost}) will give the piecewise linear form (\ref{Cform})
for the moments. If \( f(\xi )\sim \xi ^{-\beta } \) for large \( \xi  \),
than low moments \( M_{q} \) with \( q<\beta -1 \) will be determined by the
central, self-similar part of the distribution, and high moments (\( q>\beta -1 \))
by the cutoff value, 
\begin{equation}
\label{almmom}
M_{q}(t)\sim \left\{ \begin{array}{ccc}
t^{\nu q} & {\mbox {for}} & q<\beta -1\\
t^{q-(1-\nu )(\beta -1)} & {\mbox {if}} & q>\beta -1
\end{array}\right. 
\end{equation}
 which is equivalent to (\ref{Cform}) with 
\begin{equation}
\label{Cconst}
c=(1-\nu )(\beta -1)
\end{equation}
 We may conclude, that the piecewise linear dependence of the exponent \( \mu (q) \)
on the moment number \( q \) is a signature of an almost self-similar evolution
of tracer distribution with a long-tailed \( f(\xi ) \). The constant \( c \)
in (\ref{Cform}) is related to the self-similarity exponent \( \nu  \) and
power law decay exponent \( \beta  \) of \( f(\xi ) \) by (\ref{Cconst}).

Another consequence of the intermittent character of tracer motion is an anomalous
distribution of recurrences of the Poincar\'{e} map of tracer trajectories.
To define recurrences, we take a region \( B \) in the chaotic sea, and register
all returns of a Poincar\'{e} map trajectory into \( B \). The length of a
recurrence is a time interval between two successive returns. In a system with
perfect mixing, the PDF of recurrence lengths obeys a Poissonian law, provided
\( B \) is small enough, and decay of the long-recurrence tail of the distribution
is exponential for any \( B \). Recurrence distributions for tracers in all
three cases (\( k=0.2 \), \( k=0.3 \) and \( k=0.41 \)) are shown in Figures
\ref{poincarerec02}, \ref{poincarerec03}, \ref{poincarerec041}. The plots
show, that all distributions have long tails, indicating, that between the returns
tracers are being trapped in long flights of highly correlated motion. The form
of the graphs suggests, that long recurrences are distributed according to a
power law 
\begin{equation}
P(t)\sim t^{-\gamma }
\end{equation}
 The values of the exponent \( \gamma (\delta ) \) are: 
\begin{equation}
\gamma (0.2)=2.2\, \, \, \, \, \, \gamma (0.3)=2.4\, \, \, \, \, \, \gamma (0.41)=3.1\: .
\end{equation}
 Note, that while the collapse configuration is approached, the value of the
exponent increases, which may be interpreted as an improvement in mixing properties
of the flow. This agrees with the changes in the structure of Poincar\'{e}
section (Fig. \ref{specialspeedsk02}-\ref{specialspeedk041}): the closer to
collapse we get, the bigger part of the chaotic domain is occupied by a well-mixed
area, and the smaller is the role of the singular zones around KAM islands.

In fact one can try to find out the influence of the different islands on transport
by using the distributions illustrated in Figures \ref{speeddistribk02}-\ref{probdistribsk041}.
Indeed, each island corresponds to a specific peak. We recompute the moments
of the distribution in the far from collapse case \( k=0.2 \), for the modified
data set, where the trajectories, corresponding to a specific peak are discarded.
The result is presented in Fig.~\ref{expovsmomforcutk02}. We notice that the
cores do affect the transport, but their influence is essentially visible for
the high moments, while the slow particles trapped in the outer rim, are mainly
responsible for the low moments; we also notice that we do not observe the change
in slope of the strong anomalous behavior anymore, and conclude that the strong
anomalous feature is due to the interplay of the different structures in the
phase plane.

\section{Kinetics of advected particles}

In some of the previous publications (see, for example, \cite{Chernikov90}\cite{ZEN97}-\cite{ZasEdel2000})
it was clearly indicated that the properties of anomalous transport are sensitive
to phase space topology. More specifically, if we use the fractional kinetic
equation \cite{Zaslavsky92,ZEN97} in the form 
\begin{equation}
\label{equ5_{1}}
{\partial ^{\beta }P(\theta ,t)\over \partial t^{\beta }}={\mathcal{D}}{\partial ^{\alpha }P(\theta ,t)\over \partial |\theta |^{\alpha }}
\end{equation}
 to describe distributions \( P(\theta ,t) \) of rotations over angle \( \theta  \),
then the transport coefficient \( \mathcal{D} \) and exponents \( (\alpha ,\beta ) \)
depend on the presence of different structures such as boundaries of the domain,
islands, cantori, etc. The results of Section 4 show the stickiness
of trajectories of advected particles to the boundary of the domain and to boundaries
of islands. This phenomenon is similar to what has been observed in \cite{KZ98.1}
for the same-sign vortices. Our goal of this section is to estimate the values
of the exponents \( \alpha ,\beta  \).

\begin{figure}[!h]
{\par\centering \resizebox*{12cm}{!}{\includegraphics{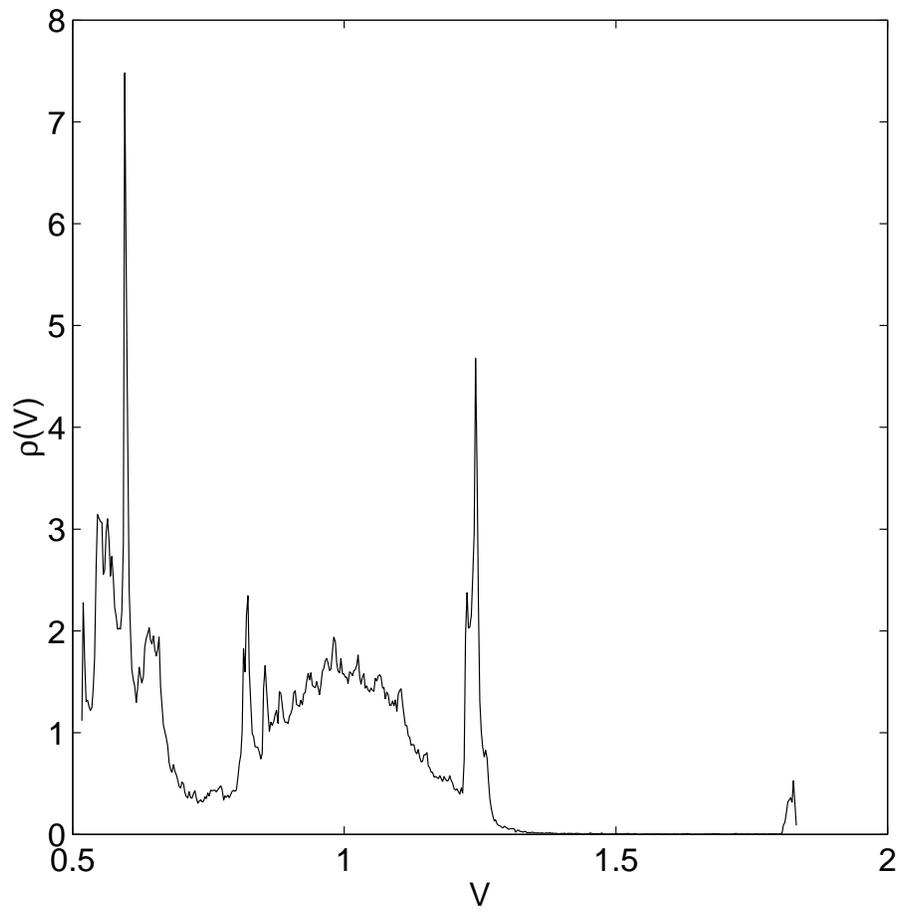}} \par}

\caption{Probability distribution of \protect\( V=(s(t+10T)-s(t))/10T\protect \), where
\protect\( s\protect \) is the arclength of a given tracer. We notice different
peaks. The constant of motion are \protect\( K=0\protect \), \protect\( \Lambda =0.9\protect \).
Vortex strengths are \protect\( (-0.2,\: 1,\: 1)\protect \). The period of
the motion is \protect\( T=10.726\protect \).\label{speeddistribk02}}
\end{figure}

\begin{figure}[!h]
{\par\centering \resizebox*{12cm}{!}{\includegraphics{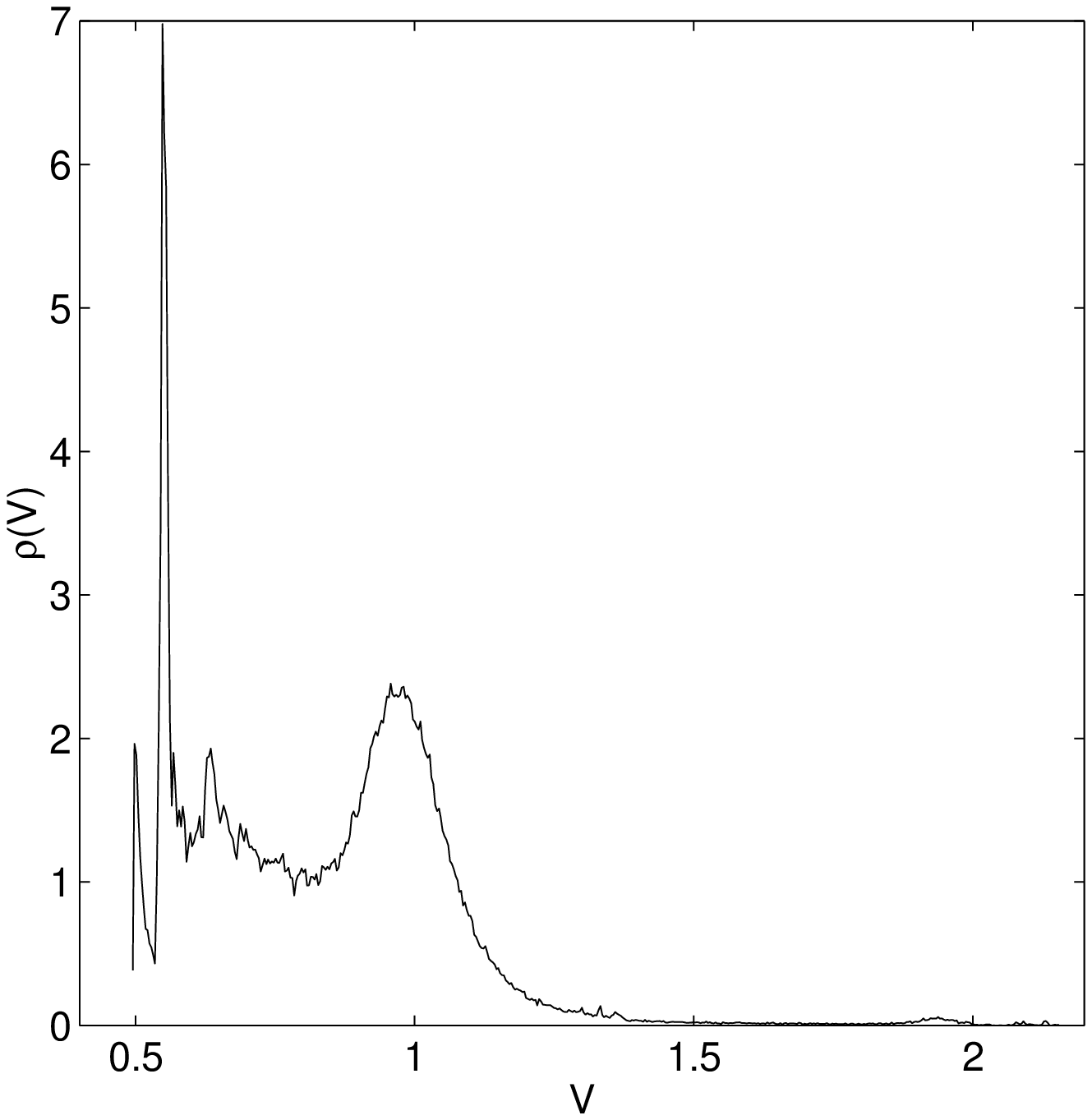}} \par}

\caption{Probability distribution of \protect\( V=(s(t+10T)-s(t))/10T\protect \). We notice different
peaks.\label{probdistribsk03} The constants of motion are \protect\( K=0\protect \),
\protect\( \Lambda =0.9\protect \). Vortex strengths are \protect\( (-0.3,\: 1,\: 1)\protect \).
The period of the motion is \protect\( T=17.53\protect \).}
\end{figure}

\begin{figure}[!h]
{\par\centering \resizebox*{12cm}{!}{\includegraphics{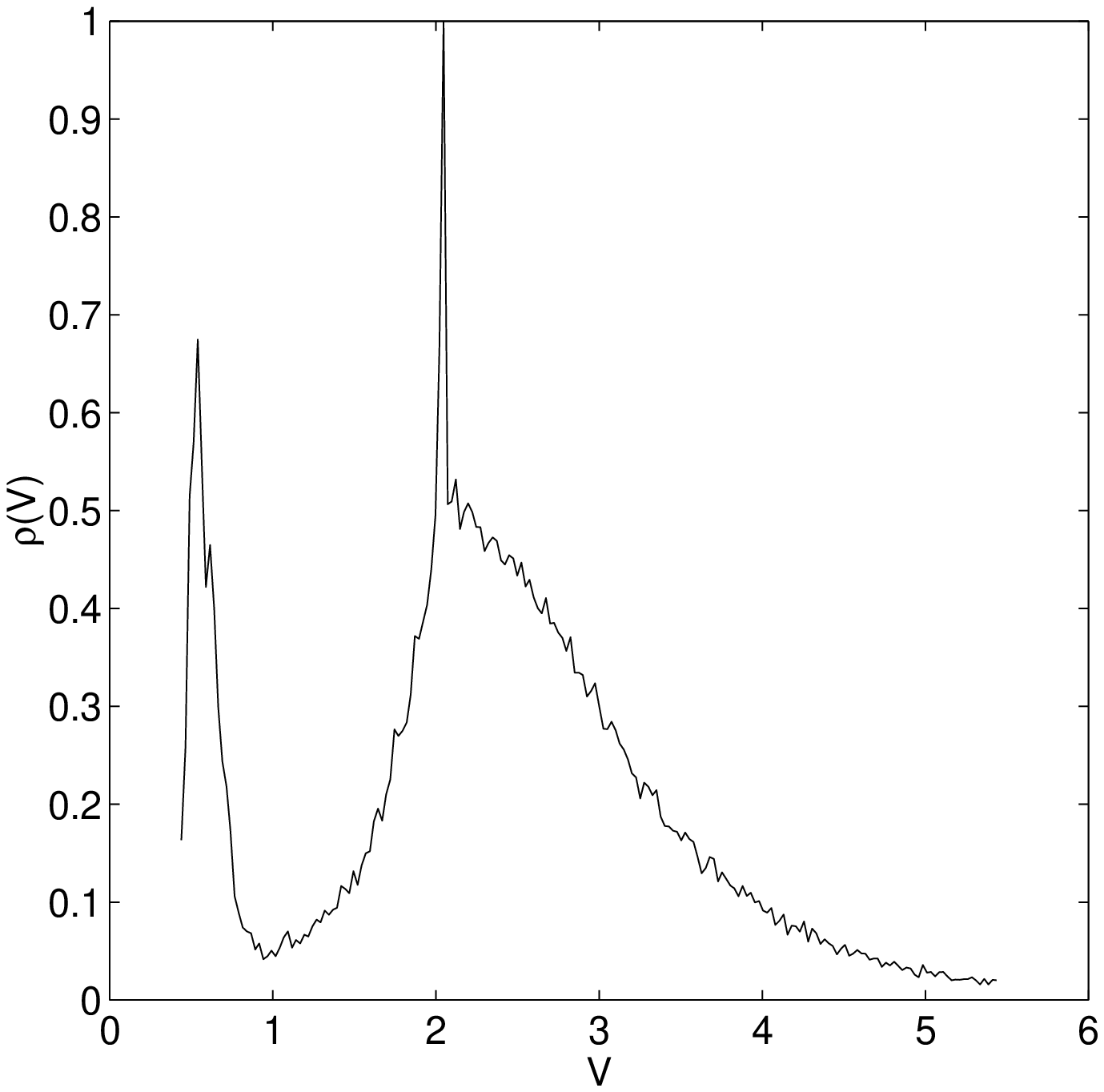}} \par}

\caption{Probability distribution of \protect\( V=(s(t+10T)-s(t))/10T\protect \). We notice different
peaks.\label{probdistribsk041} The constants of motion are \protect\( K=0\protect \),
\protect\( \Lambda =0.9\protect \). Vortex strengths are \protect\( (-0.41,\: 1,\: 1)\protect \).
The period of the motion is \protect\( T=36.86\protect \).}
\end{figure}

\begin{figure}[!h]
{\par\centering \resizebox*{12cm}{!}{\includegraphics{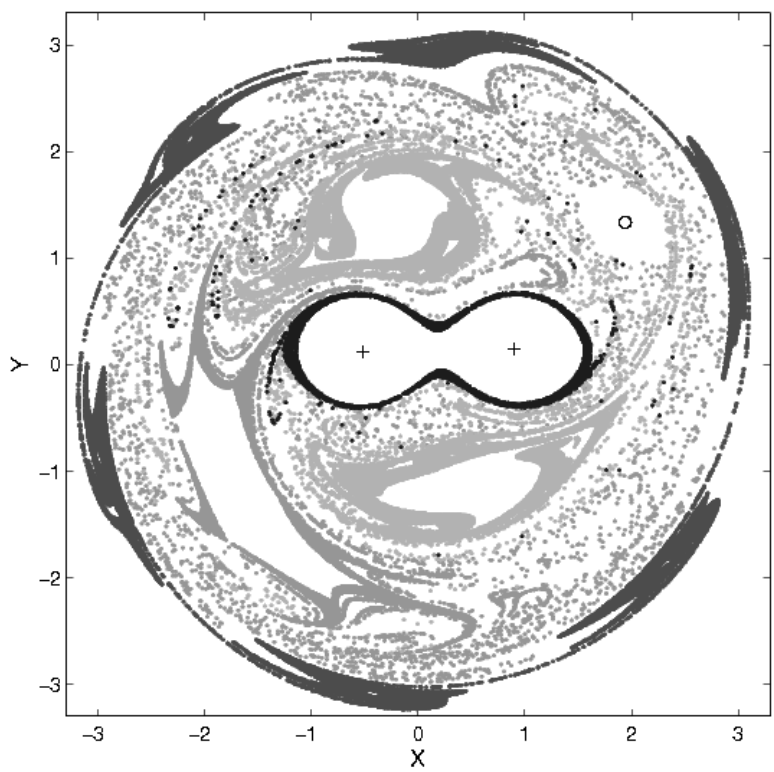}} \par}

\caption{Location on the Poincar\'e section of the points with definite ``special''
averaged velocity (compare to peaks in Fig. \ref{speeddistribk02}).\label{specialspeedsk02} In dark gray the points corresponding
to \protect\( 0.595<V<0.605\protect \). In light and lighter gray the points corresponding respectively to
\protect\( 0.81<V<0.83\protect \) and to \protect\( 1.22<V<1.25\protect \).
In black the points with corresponding speeds \protect\( V>1.8\protect \). The
constants of motion are \protect\( K=0\protect \), \protect\( \Lambda =0.9\protect \).
Vortex strengths are \protect\( (-0.2,\: 1,\: 1)\protect \). The period of
the motion is \protect\( T=10.73\protect \).}
\end{figure}

\begin{figure}[!h]
{\par\centering \resizebox*{12cm}{!}{\includegraphics{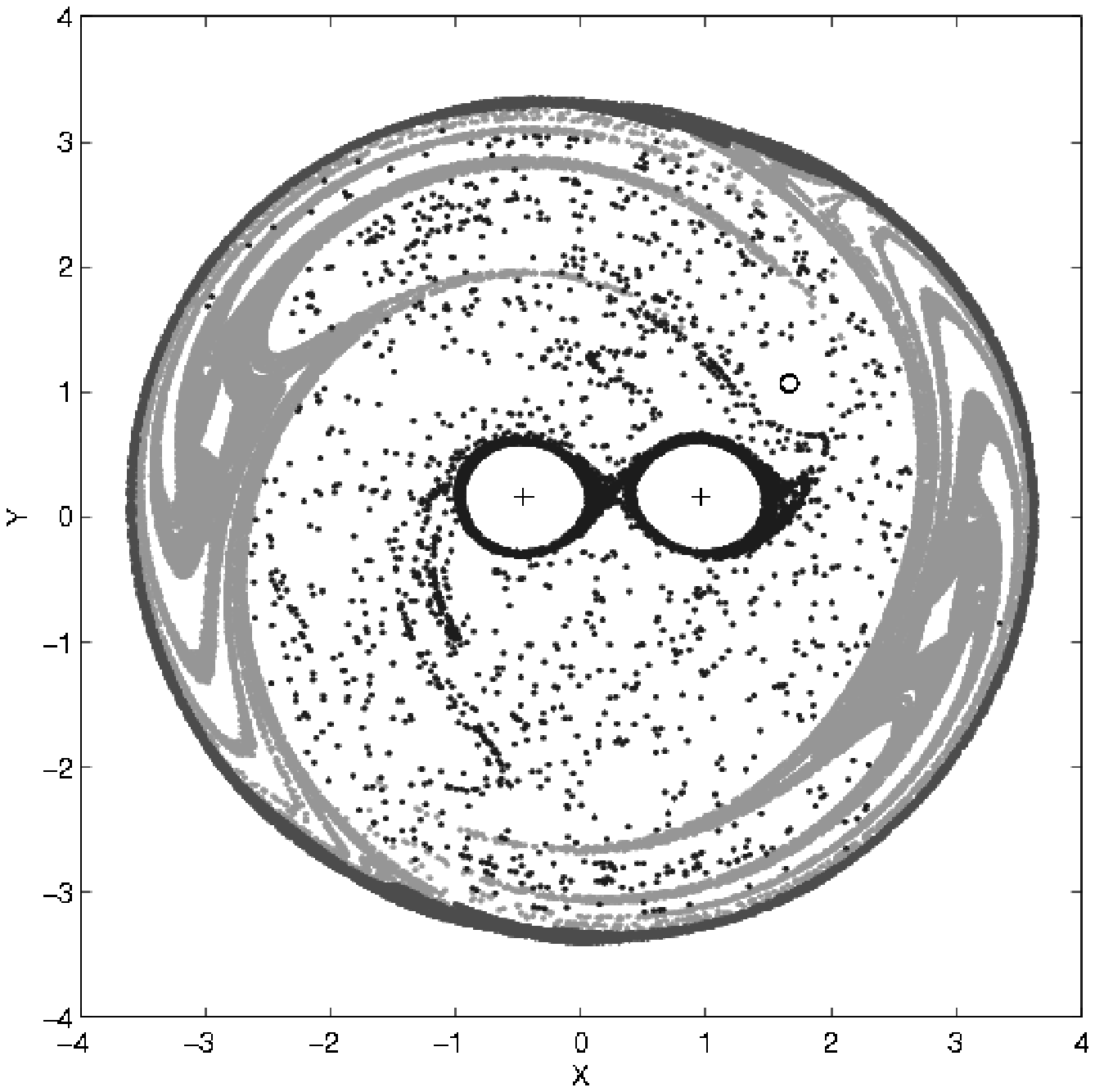}} \par}

\caption{Location on the Poincar\'e section of the points with definite ``special''
velocity (compare to peaks in Fig. \ref{probdistribsk03}).\label{specialspeedk03} In light gray the points corresponding to \protect\( 0.545<V<0.565\protect \).
In black the points with corresponding speeds \protect\( V>1.5\protect \). In
dark gray the points corresponding to \protect\( V<0.52\protect \). The constants
of motion are \protect\( K=0\protect \), \protect\( \Lambda =0.9\protect \).
Vortex strengths are \protect\( (-0.3,\: 1,\: 1)\protect \). The period of
the motion is \protect\( T=17.53\protect \).}
\end{figure}

\begin{figure}[!h]
{\par\centering \resizebox*{12cm}{!}{\includegraphics{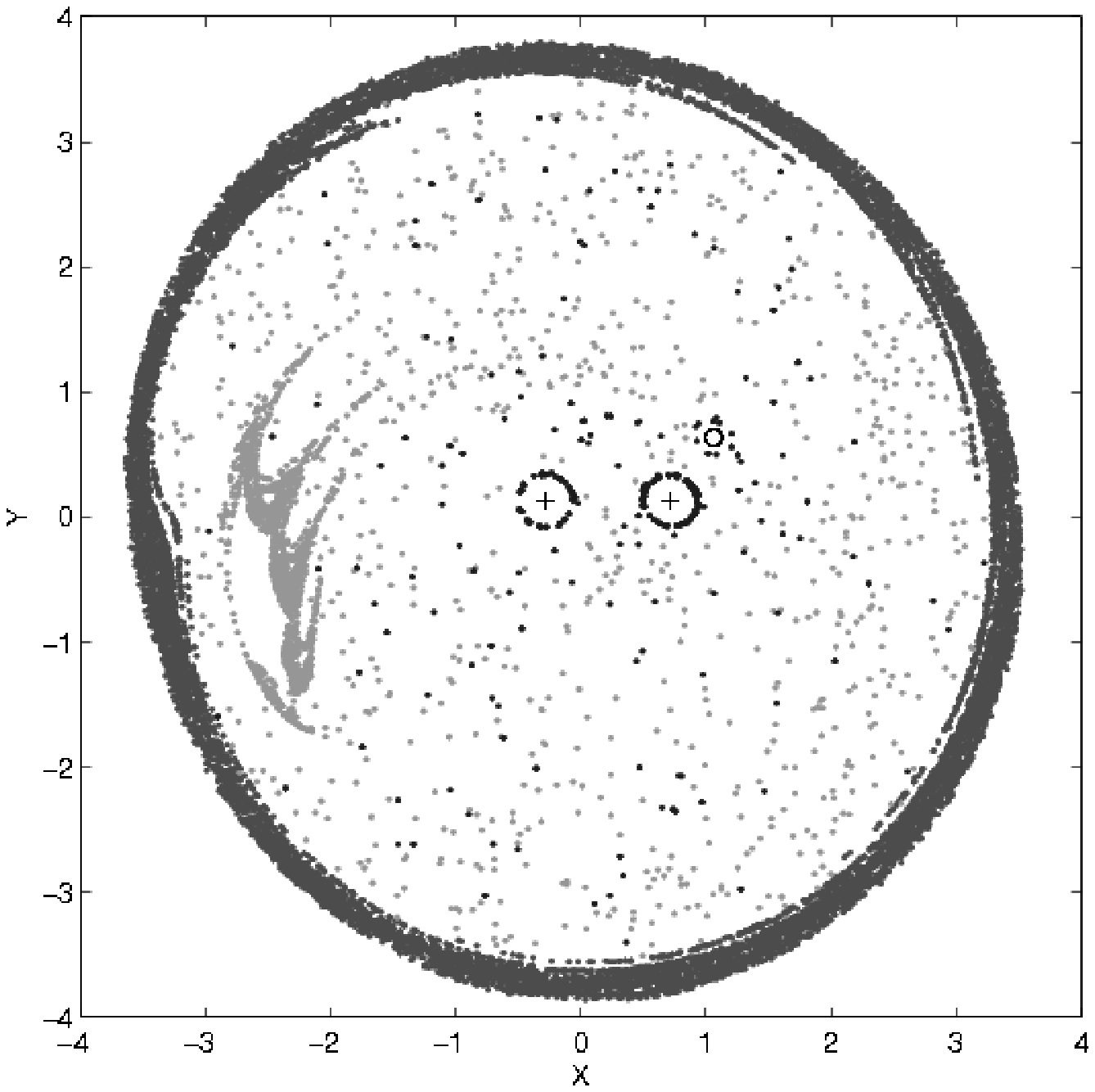}} \par}

\caption{Location on the Poincar\'e section of the points with definite ``special''
angular velocity (compare to peaks in Fig. \ref{probdistribsk041}).\label{specialspeedk041} In light gray the points corresponding
to \protect\( 0.7488<V<0.7515\protect \). In black the points with corresponding
speeds \protect\( V>1.5\protect \). In dark gray the points corresponding to \protect\( V<0.489\protect \)
The constants of motion are \protect\( K=0\protect \), \protect\( \Lambda =0.9\protect \).
Vortex strengths are \protect\( (-0.41,\: 1,\: 1)\protect \). The period of
the motion is \protect\( T=36.86\protect \).}
\end{figure}

\begin{figure}[!h]
{\par\centering \resizebox*{12cm}{!}{\includegraphics{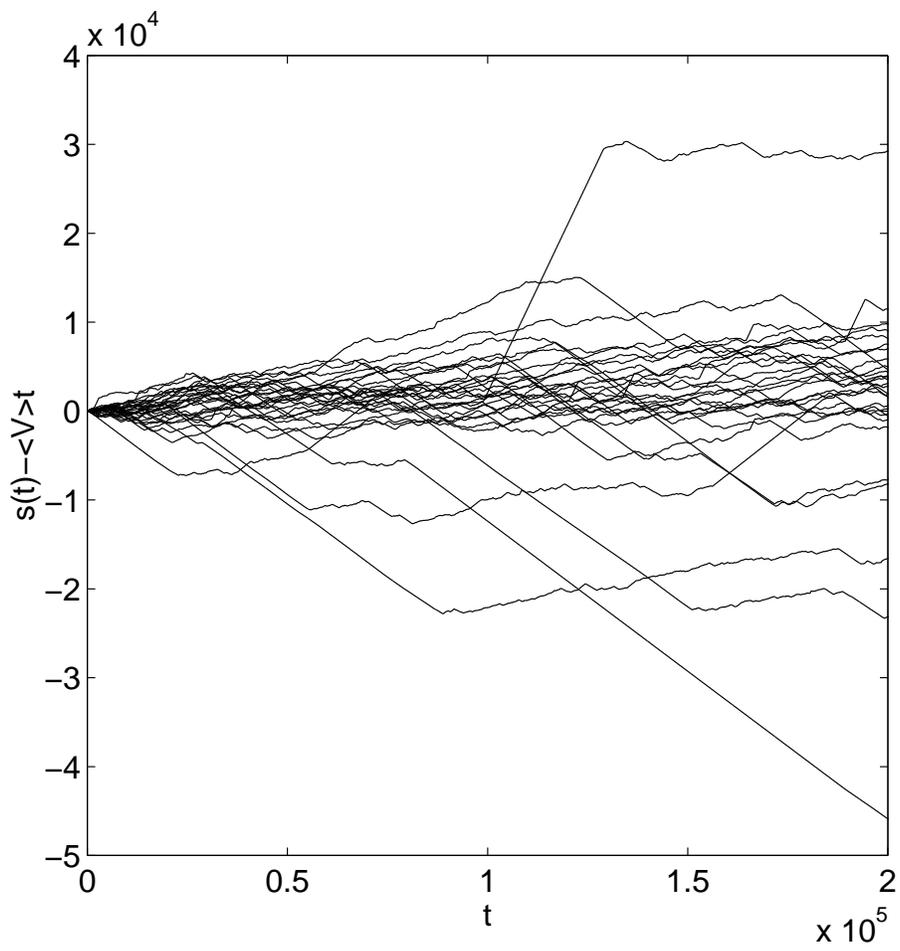}} \par}

\caption{The deviation from the mean arclength (\protect\( s(t)-Vt\protect \)) versus
time \protect\( t\protect \) is plotted for \protect\( 30\protect \) particles.
We notice the flights corresponding to a particle being in a the sticky zone
around an island. The constants of motion are \protect\( K=0\protect \), \protect\( \Lambda =0.9\protect \).
The run is over \protect\( 20000\protect \) periods. The average speed is \protect\( V\approx 0.87\protect \).
Vortex strengths are \protect\( (-0.2,\: 1,\: 1)\protect \). The period of
the motion is \protect\( T=10.73\protect \).\label{deviationfrommeank02}}
\end{figure}

\begin{figure}[!h]
{\par\centering \resizebox*{12cm}{!}{\includegraphics{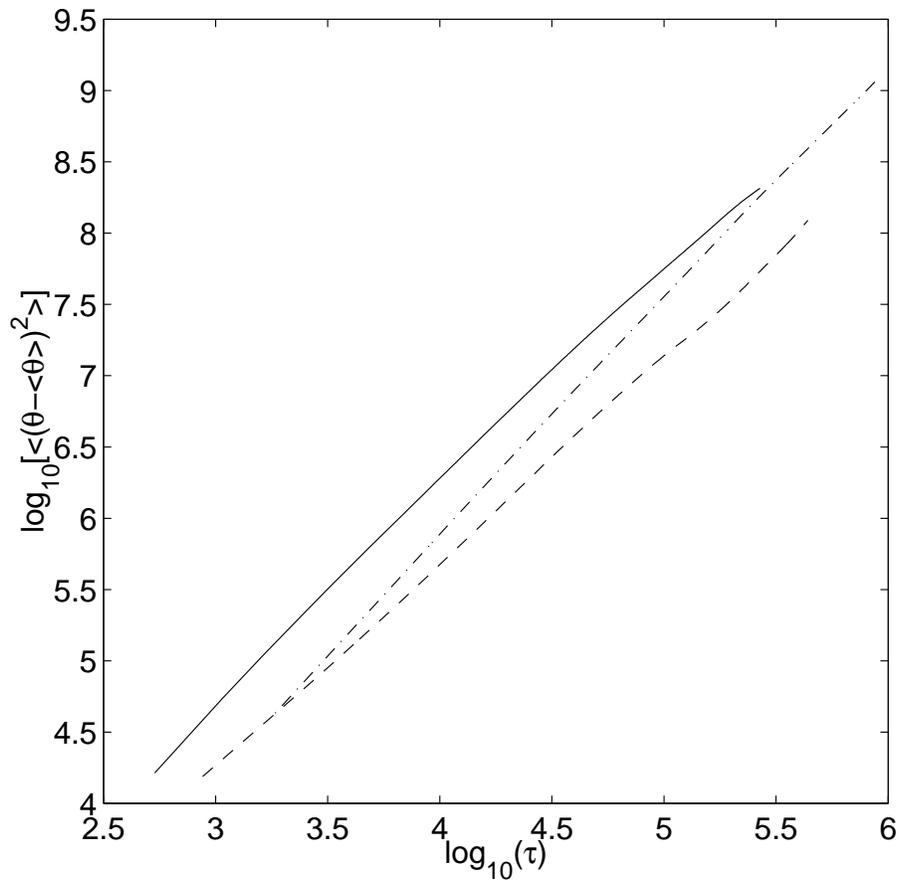}} \par}

\caption{The second order moment for the angle distribution 
(\protect\( \langle (\theta (t)-\overline{\theta }(t))^{2}\rangle \protect \))
versus time for the three cases \protect\( k=0.2\protect \) (solid line), \protect\( k=0.3\protect \)
(dashed line), and \protect\( k=0.41\protect \) (dot-dashed line). We notice
a change of behavior for the large times. The constants of motion are \protect\( K=0\protect \),
\protect\( \Lambda =0.9\protect \). \label{figsecondmomentvstime}}
\end{figure}

\begin{figure}[!h]
{\par\centering \resizebox*{12cm}{!}{\includegraphics{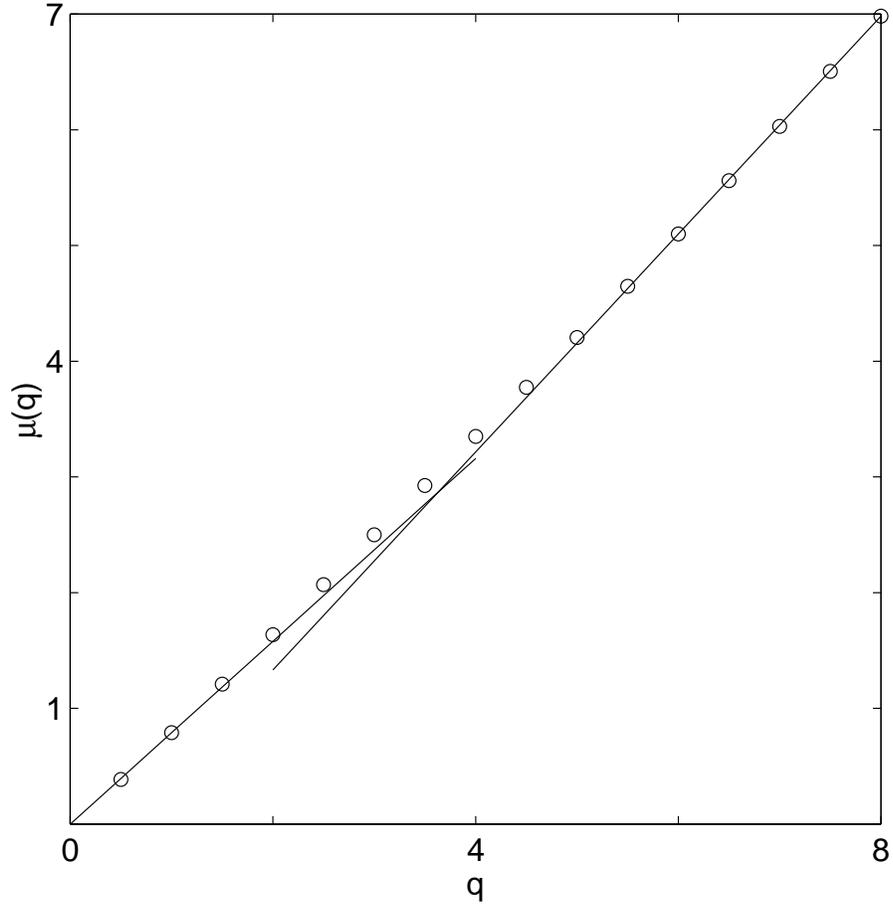}} \par}

\caption{The exponent $q\nu (q)$ versus the moment order  $q$ 
for the angle distribution
($\langle |\theta (t)-\overline{\theta }(t)|^{q}\rangle \sim t^{q\nu (q)}$)
is plotted for the small times ($ t<3\; 10^{4}$, or $ t<3000T$).
We notice two linear behaviors: 
$\mu (q)=  0.79q$ for $q<2$, and
$\mu (q)=  0.94q+Cte$ for $q>4$
Vortex strengths are $ (-0.2,\: 1,\: 1)$. The period of
the motion is $T=10.7$.\label{exponentssmalltimesk02}}
\end{figure}

\begin{figure}[!h]
{\par\centering \resizebox*{12cm}{!}{\includegraphics{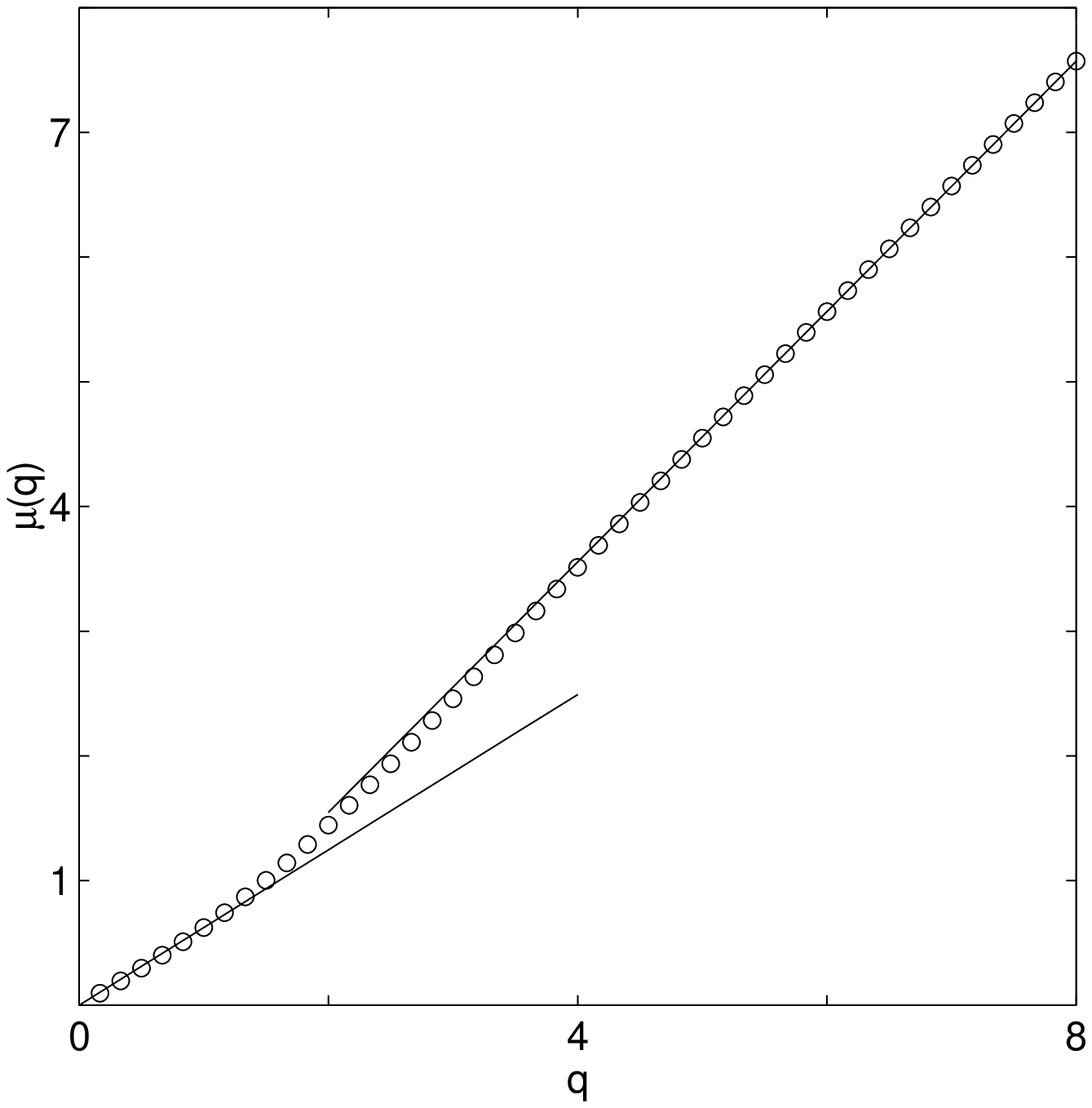}} \par}

\caption{The exponent \protect\( \mu (q)\protect \) versus the moment order \protect\( q\protect \)
for the angle distribution (\protect\( \langle |\theta (t)-\overline{\theta }(t)|^{q}\rangle \sim t^{\mu (q)}\protect \))
is plotted for the small times (\protect\( t<5\; 10^{4}\protect \), or \protect\( t<3000T\protect \))\label{Figexpvsmomsmalltk03}.
We notice two linear behaviors:
$ \mu (q)=  0.62q$ ($q<2$),
$\mu (q)=  1.00q+Cte$ ($q>2$).
 The constants of motion are \protect\( K=0\protect \), \protect\( \Lambda =0.9\protect \).
Vortex strengths are \protect\( (-0.3,\: 1,\: 1)\protect \). The period of
the motion is \protect\( T=17.53\protect \).}
\end{figure}

\begin{figure}[!h]
{\par\centering \resizebox*{12cm}{!}{\includegraphics{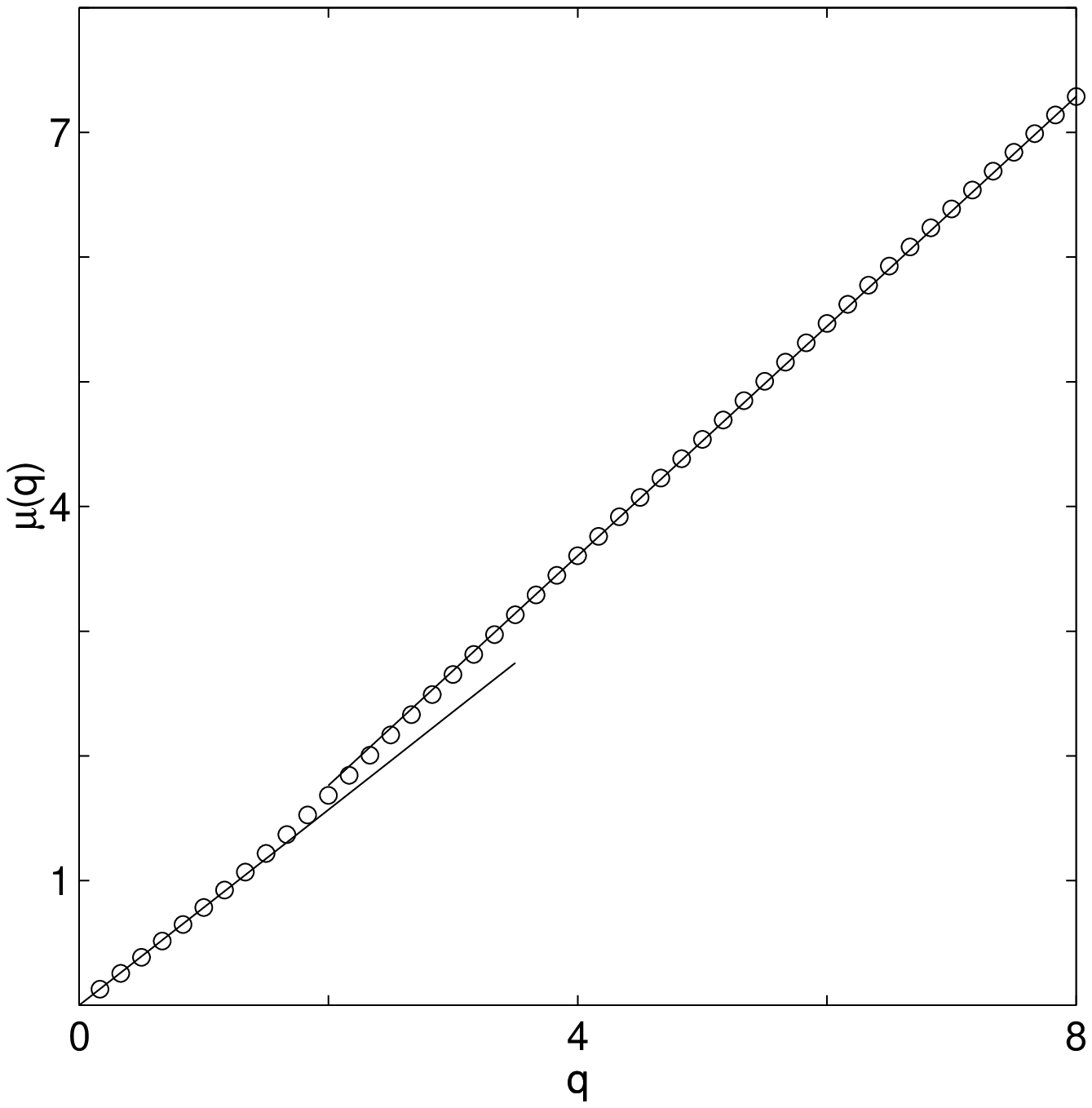}} \par}

\caption{The exponent \protect\( \mu (q)\protect \) versus the moment order \protect\( q\protect \)
for the angle distribution (\protect\( \langle |\theta (t)-\overline{\theta }(t)|^{q}\rangle \sim t^{\mu (q)}\protect \))
is plotted for the short times (\protect\( t<10^{5}\protect \))\label{Figexpvsmomsmaltk041}.
We notice two linear behaviors:
 $\mu (q)=  0.78q$ ($q<2$),
 $ \mu (q)=  0.92q-Cte$ ($q>2$).
 The constants of motion are \protect\( K=0\protect \), \protect\( \Lambda =0.9\protect \).
Vortex strengths are \protect\( (-0.41,\: 1,\: 1)\protect \). The period of
the motion is \protect\( T=36.85\protect \).}
\end{figure}

\begin{figure}[!h]
{\par\centering \resizebox*{12cm}{!}{\includegraphics{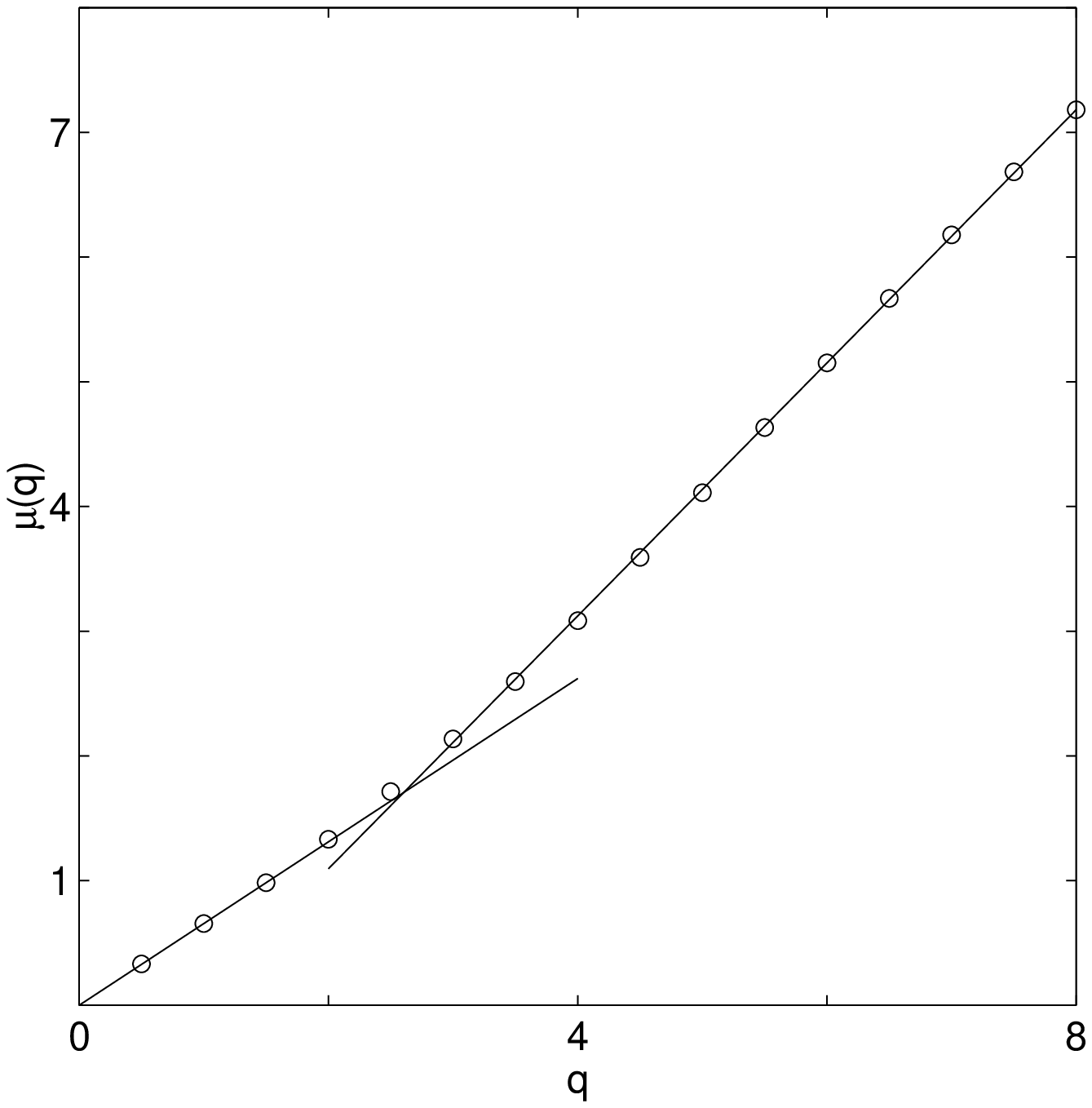}} \par}

\caption{The exponent \protect\( \mu (q)\protect \) versus the moment order \protect\( q\protect \)
\label{exponentslongtimesk02}
for the angle distribution (\protect\( \langle |\theta (t)-\overline{\theta }(t)|^{q}\rangle \sim t^{\mu (q)}\protect \))
is plotted for the long times (\protect\( t>10^{5}\protect \)). We notice two
linear behaviors: 
$ \mu (q)=  0.62q$ ($q<2$),
$\mu (q)=  1.01q-Cte$ ($q>2$).
 The constants of motion are \protect\( K=0\protect \), \protect\( \Lambda =0.9\protect \).
Vortex strengths are \protect\( (-0.2,\: 1,\: 1)\protect \). The period of
the motion is \protect\( T=10.7\protect \).}
\end{figure}

\begin{figure}[!h]
{\par\centering \resizebox*{12cm}{!}{\includegraphics{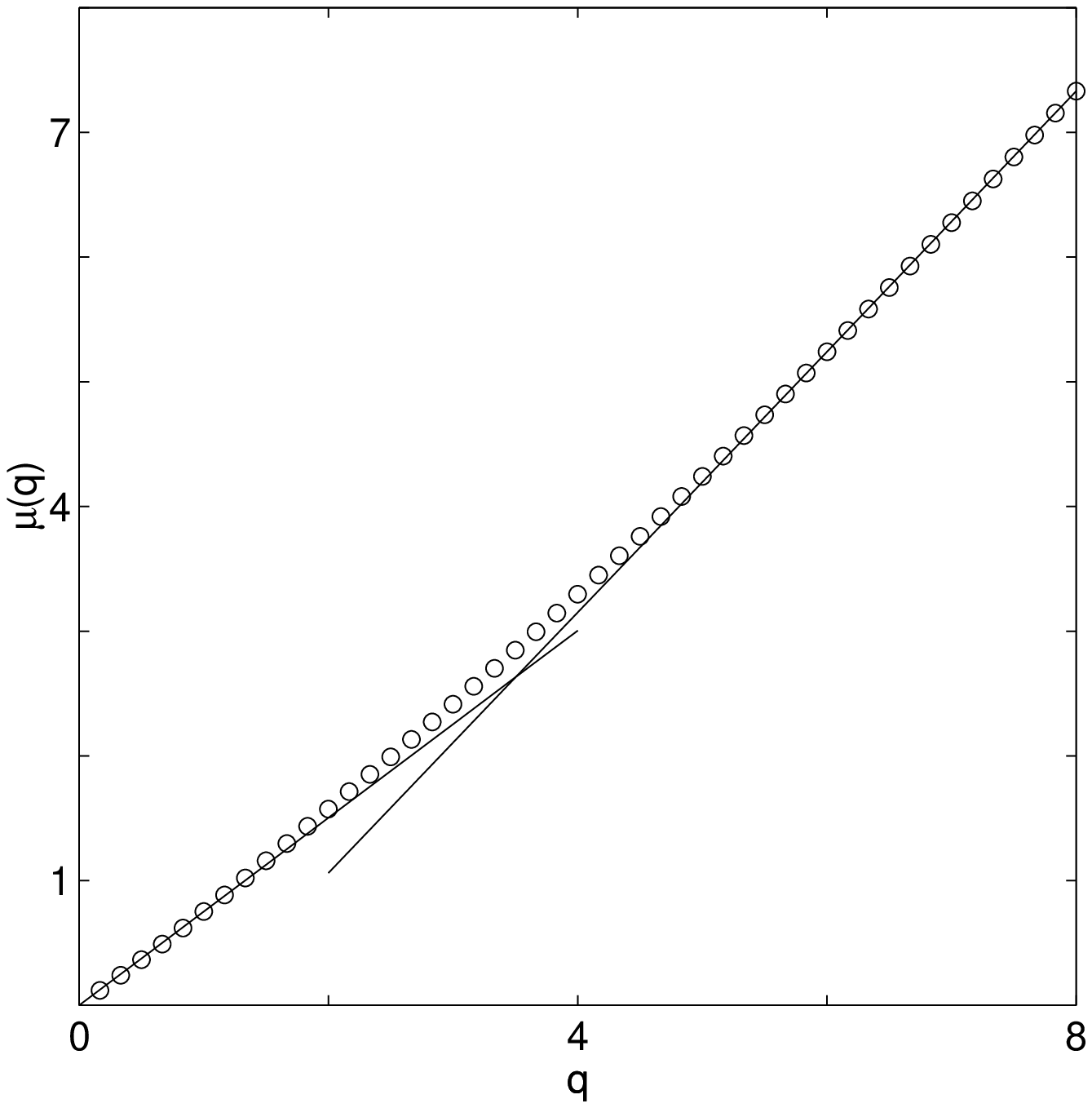}} \par}

\caption{The exponent \protect\( \mu (q)\protect \) versus the moment order \protect\( q\protect \)
\label{Figexpvsmomlongtk03}
for the angle distribution (\protect\( \langle |\theta (t)-\overline{\theta }(t)|^{q}\rangle \sim t^{\mu (q)}\protect \))
is plotted for the long times (\protect\( t>1.5\; 10^{5}\protect \)). We notice
two linear behaviors:
$\mu (q)=  0.75q$ ($q<2$),
$\mu (q)=  1.04q-Cte$ ($q>2$).
 The constants of motion are \protect\( K=0\protect \), \protect\( \Lambda =0.9\protect \).
Vortex strengths are \protect\( (-0.3,\: 1,\: 1)\protect \). The period of
the motion is \protect\( T=17.53\protect \).}
\end{figure}

\begin{figure}[!h]
{\par\centering \resizebox*{12cm}{!}{\includegraphics{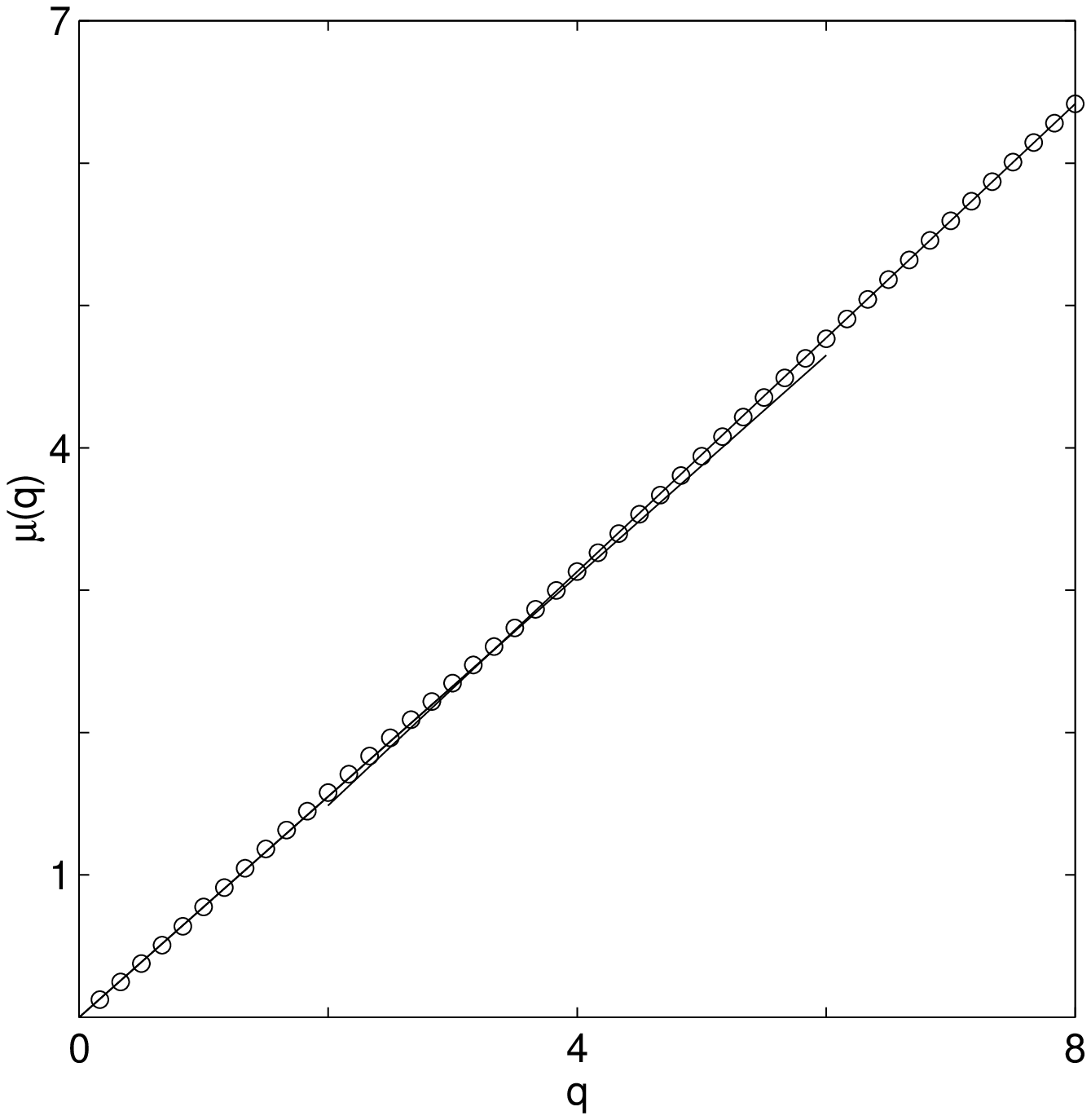}} \par}

\caption{The exponent \protect\( \mu (q)\protect \) versus the moment order \protect\( q\protect \)
\label{Figexpvsmomlongtk041}
for the angle distribution (\protect\( \langle |\theta (t)-\overline{\theta }(t)|^{q}\rangle \sim t^{\mu (q)}\protect \))
is plotted for the long times (\protect\( 10^{6}>t>5\; 10^{5}\protect \)).
We notice two linear behaviors:
$ \mu (q)=  0.77q$ ($q<2$),
$ \mu (q)=  0.82q-Cte$ ($q>2$).
 The constants of motion are \protect\( K=0\protect \), \protect\( \Lambda =0.9\protect \).
Vortex strengths are \protect\( (-0.41,\: 1,\: 1)\protect \). The period of
the motion is \protect\( T=36.85\protect \).}
\end{figure}

\begin{figure}[!h]
{\par\centering \resizebox*{12cm}{!}{\includegraphics{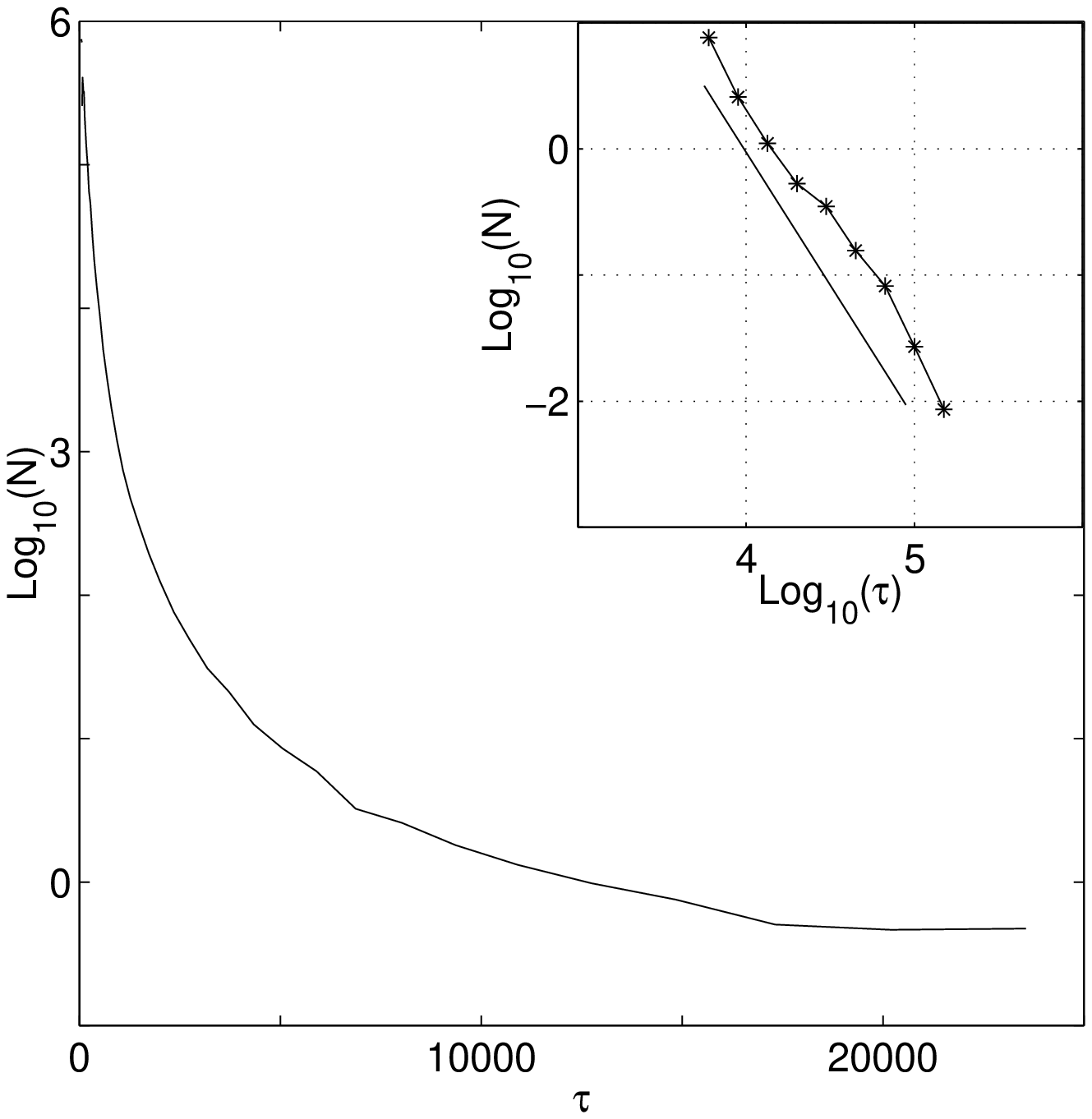}} \par}

\caption{Distribution of Poincar\'e recurrences. The constants of motion are \protect\( K=0\protect \),
\protect\( \Lambda =0.9\protect \). Vortex strengths are \protect\( (-0.2,\: 1,\: 1)\protect \).
The period of the motion is \protect\( T=10.7\protect \). The tail presents
a power law behavior whose exponent is \protect\( \sim 2.2\protect \). The
simulation is performed over 50 000 periods, statistics are made with 1137 particles.\label{poincarerec02}}
\end{figure}

\begin{figure}[!h]
{\par\centering \resizebox*{12cm}{!}{\includegraphics{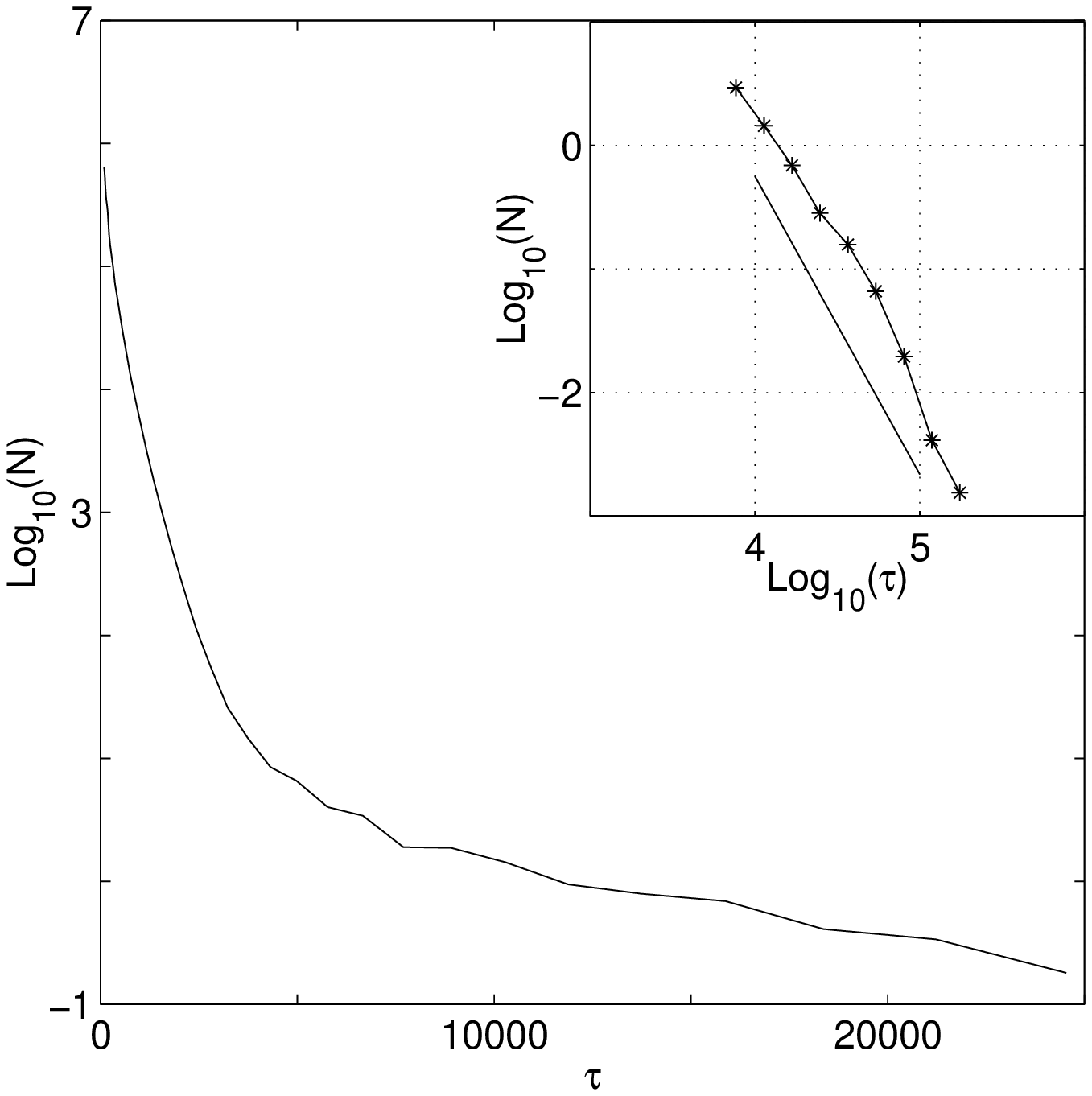}} \par}

\caption{Distribution of Poincar\'e recurrences. The constants of motion are \protect\( K=0\protect \),
\protect\( \Lambda =0.9\protect \). Vortex strengths are \protect\( (-0.3,\: 1,\: 1)\protect \).
The period of the motion is \protect\( T=17.53\protect \). The tail presents
a power law behavior whose exponent is \protect\( \sim 2.4\protect \). The
simulation is performed over 50 000 periods, statistics are made with 1012 particles.\label{poincarerec03}}
\end{figure}

\begin{figure}[!h]
{\par\centering \resizebox*{12cm}{!}{\includegraphics{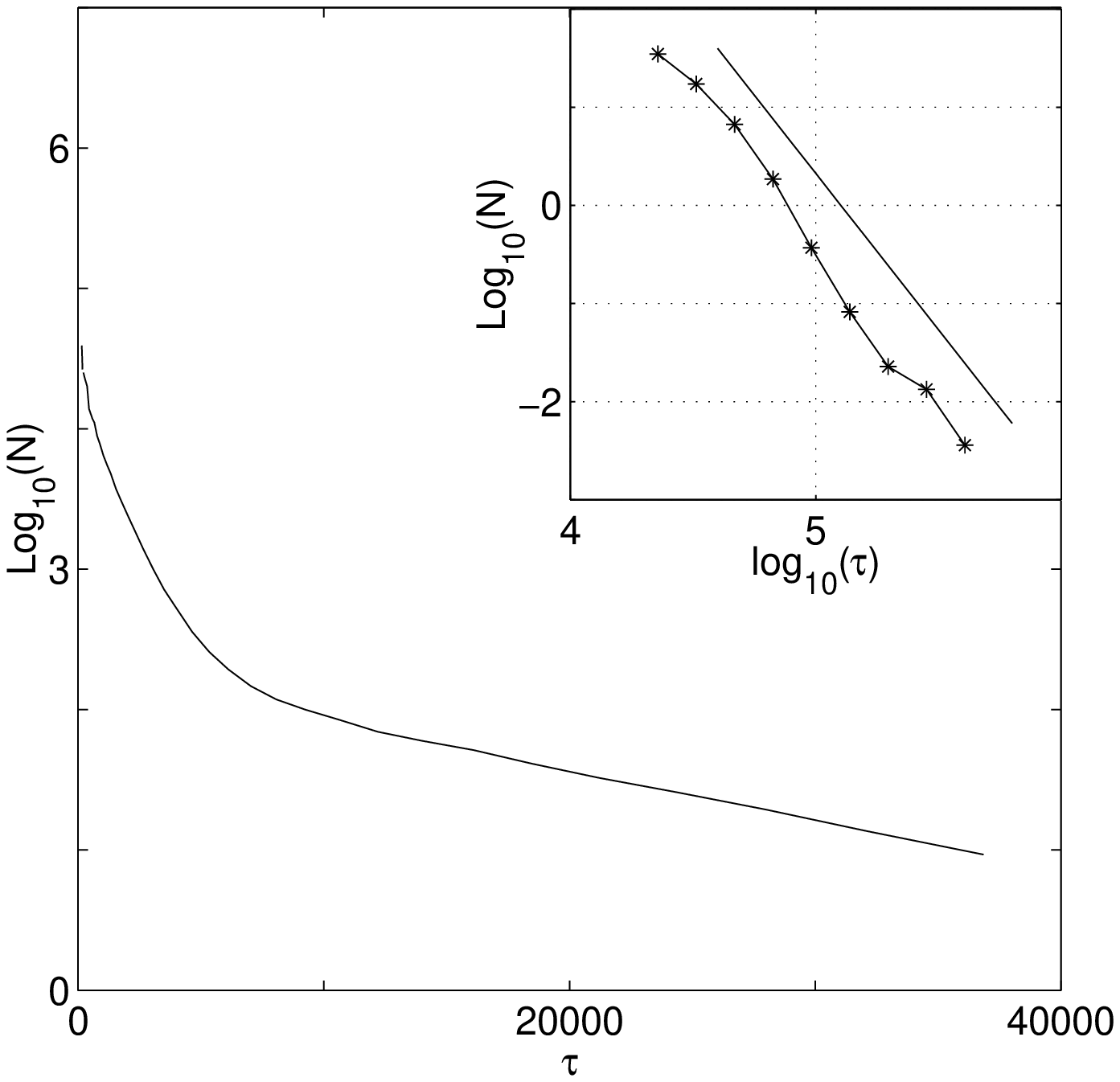}} \par}

\caption{Distribution of Poincar\'e recurrences. The constants of motion are \protect\( K=0\protect \),
\protect\( \Lambda =0.9\protect \). Vortex strengths are \protect\( (-0.41,\: 1,\: 1)\protect \).
The period of the motion is \protect\( T=36.85\protect \). The tail presents
a power law behavior whose exponent is \protect\( \sim 3.1\protect \). The
simulation is performed over 50 000 periods, statistics are made with 637 particles.\label{poincarerec041}}
\end{figure}

\begin{figure}[!h]
{\par\centering \resizebox*{12cm}{!}{\includegraphics{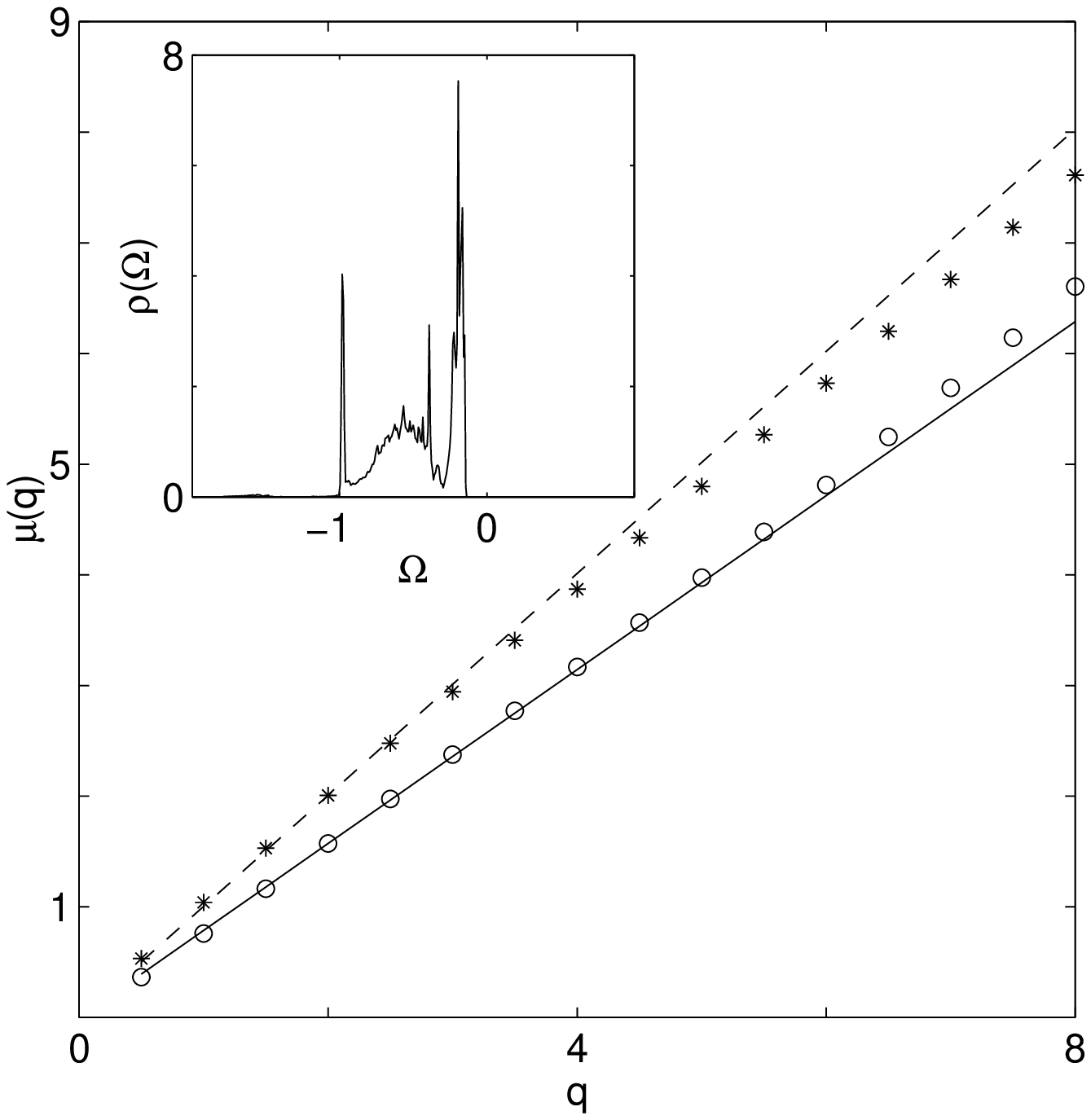}} \par}

\caption{Exponents versus moment order in the case \protect\( k=0.2\protect \), the
difference with Fig.\ref{exponentssmalltimesk02} or Fig.\ref{exponentslongtimesk02}
is that here we did cut some parts of the distribution. the star sign ``{*}''
the slow motion (near zero peak corresponding to the outer rim) has been removed, the circle sign ``o'' the
fast motion ($\Omega<-1$ corresponding to the cores) has been removed. We notice two linear behavior 
with slopes $0.78$ and $1$, which is respectively similar to the low and
high moment behavior of  Fig.\ref{exponentssmalltimesk02} or Fig.\ref{exponentslongtimesk02}.
This allows us to identify the individual role in transport of the different structures in the phase
space.
\label{expovsmomforcutk02}}
\end{figure}

Figures \ref{specialspeedsk02}-\ref{specialspeedk041} demonstrate stickiness
of trajectories to specific structures with a filamentation of sticky domains
along stable/unstable manifolds. In fact, different sticky domains generate
different intermittent scenarios with some associated values of \( (\alpha ,\beta ) \)
\cite{ZEN97,ZasEdel2000}. As a result, the real kinetics is multi-fractional
and can be characterized by a set of values of \( (\alpha ,\beta ) \) or, more
precisely, by a spectral function of \( (\alpha ,\beta ) \) in the same sense
as the spectral function for multi-fractals \cite{Hentschel84}-\cite{Jensen85}.
Figures \ref{speeddistribk02}-\ref{probdistribsk041} show that trajectories,
sticking to different structures (islands), have different angular velocities
(compare to peaks in Figures \ref{speeddistribk02}-\ref{probdistribsk041}).
Due to this, different asymptotics to the distribution function \( P(\theta ,t) \)
and different values of \( (\alpha ,\beta ) \) will appear for different time
intervals. In other words, for a considered time interval one can expect a specific
``intermediate asymptotics'' for \( P(\theta ,t) \) and, correspondingly,
different pairs \( (\alpha ,\beta ) \). Different classes of universality for
the values \( (\alpha ,\beta ) \) were discussed in \cite{ZasEdel2000}. Below
we will apply some of these results. 
\begin{table}
{\centering \begin{tabular}{|c|c|c|}
\hline 
\( k \) &
 \( q=2 \)&
 \( q>2 \)\\
\hline 
\( 0.2 \)&
 1.6 &
 1.88 \\
\hline 
 \( 0.3 \)&
 1.4 &
 2.0 \\
\hline 
 \( 0.41 \)&
 1.6 &
 1.84  \\
\hline 
\end{tabular}\par}

\caption{Values of the transport exponents \protect\( \mu \protect \) for different
moments \protect\( q\protect \). \label{tablesec5}}
\end{table}

Multiplying (\ref{equ5_{1}}) by \( |\theta |^{\alpha } \) and integrating
it over \( |\theta | \) we obtain 
\begin{equation}
\label{equ5_{2}}
\langle |\theta |^{\alpha }\rangle \sim t^{\beta }
\end{equation}
 or, in the case of self-similarity the transport exponent \( \mu  \) from
the equation 
\begin{equation}
\label{equ5_{3}}
\langle |\theta |^{2}\rangle \sim t^{\mu }
\end{equation}
 can be estimated as 
\begin{equation}
\label{equ5_{4}}
\mu =2\beta /\alpha 
\end{equation}
 Expression (\ref{equ5_{3}}) should be considered with some reservations since
the second and higher moment may diverge. For a finite time \( t<t_{\textrm{max}} \)
particles reach a distance (angular rotation) \( \theta <\theta _{max} \),
which makes all moments finite. Typically all 
\begin{equation}
\label{equ5_{5}}
\theta _{max}=\omega _{max}t
\end{equation}
 and \( \omega _{max} \) (maximal angular velocity) can be reached only at
the boundary of the domain of chaotic motion (see Figs. 6-8).

Using notations (\ref{weak}), (\ref{strong}), and 
\begin{equation}
\label{equ5_{6}}
\mu (2)\equiv \mu 
\end{equation}
 we can present the results for the transport exponents \( \mu  \) in Table
\ref{tablesec5}. They are almost the same independently of how far is the control
parameter \( k \) from its critical value \( k_{c}=1/2 \) (the collapse condition).
For large values of \( q \) we have \( \mu  \) close to \( \mu =2 \) which
corresponds to ballistic dynamics with \( \alpha \approx \beta \approx 1 \).

This result can be well understood from the stickiness of trajectories to the
cores (see Figs. \ref{specialspeedsk02}-\ref{specialspeedk041} in the black
color). As it follows from distributions in Figs. \ref{speeddistribk02}-\ref{probdistribsk041},
the particles that stick to the cores are the fastest ones, and they just define
the large moments values.

The value of \( \mu  \) for \( q=2 \) is defined mainly by meso-structures
in the middle of Figs. \ref{specialspeedsk02}-\ref{specialspeedk041} (light gray).
 A typical property of these structures is existence of islands
with well resolved filamentations due to the vicinity of the structures to a
bifurcation. The latter is evident from the sharp corners of islands, which
may indicate a parabolic type singular point \cite{Melnikov96}. A corresponding
effective Hamiltonian, describing dynamics near a singular point, has a form
\cite{Karney83,Melnikov96,ZEN97,Rom-Kedar99}: 
\begin{equation}
\label{equ5_{7}}
H_{\textrm{eff}}=a_{1}(\Delta P)^{2}+a_{2}\Delta Q-Q_{3}(\Delta Q)^{3}
\end{equation}
 where \( (P,Q) \) are generalized momentum and coordinate and \( (\Delta P,\Delta Q) \)
are their corresponding deviations from the singular point \( (P_{0},Q_{0}) \):
\begin{equation}
\label{equ5_{8}}
\Delta P=P-P_{0}\, \, ,\, \, \, \, \, \, \, \, \Delta Q=Q-Q_{0}\, \, .
\end{equation}
 Particularly, it may be 
\begin{equation}
\label{equ5_{9}}
Q=\theta \, \, ,\, \, \, \, \, \, \, \, \, \, \, \, P=\dot{\theta }\, \, .
\end{equation}

Depending on the coefficients \( a_{j} \) and on the meaning of variables \( (P,Q) \),
which may be different from (\ref{equ5_{9}}), one can describe singularity
due to bifurcations for different types of dynamical modes: accelerator mode
\cite{Karney83,Melnikov96}, ``blinking island'' mode \cite{Melnikov96},
ballistic mode \cite{Rom-Kedar99}, etc. For all these situations, universality
of the Hamiltonian (\ref{equ5_{7}}) permits estimation of the exponents \( (\alpha ,\beta ) \)
in (\ref{equ5_{1}})

A trajectory that approaches the vicinity of the singular point (or, simply,
a corner of the island boundary), behaves intermittently and escapes the near-separatrix
boundary layer. The phase volume of the escaping trajectories is 
\begin{equation}
\label{equ5_{1}0}
\delta \Gamma =\delta P\delta Q
\end{equation}
 where \( \delta P,\delta Q \) are values \( \Delta P,\Delta Q \) related
to the escaping particles. From (\ref{equ5_{7}}) we can estimate 
\begin{equation}
\label{equ5_{1}1}
\delta P_{\textrm{max}}\sim \delta Q^{3/2}
\end{equation}
 and from (\ref{equ5_{1}0}), (\ref{equ5_{1}1}) 
\begin{equation}
\label{equ5_{1}2}
\delta \Gamma \sim \delta Q^{5/2}
\end{equation}
 Escaping from the boundary layer means growth of the ``radial'' variable
\( \delta Q \) with time, i.e. for an initial time interval \( \delta Q\sim t \),
and consequently, 
\begin{equation}
\label{equ5_{1}3}
\delta \Gamma \sim t^{5/2}\: .
\end{equation}
 From (\ref{equ5_{1}3}) we conclude for the escape probability density to leave
the boundary layer at time instant \( t \) within interval \( dt \): 
\begin{equation}
\label{equ5_{1}4}
\psi (t)\propto 1/\delta \Gamma \sim t^{-5/2}\, \, .
\end{equation}

It was shown in \cite{Montroll84} that under special conditions the exponent
\( \gamma  \) for the trapping time asymptotic distribution 
\begin{equation}
\label{equ5_{1}5}
\psi (t)\sim t^{-\gamma }
\end{equation}
 can be linked to fractal time dimension. Moreover, \( \gamma  \) is related
to the kinetic equation (\ref{equ5_{1}}) as \cite{ZEN97}
\begin{equation}
\label{equ5_{1}6}
\beta =\gamma -1\, \, .
\end{equation}
 For the considered case we have \( \beta =3/2 \).

For the spatial distribution of particles, the simplest situation occurs when
the diffusion process has Gaussian type and, consequently, \( \alpha =2 \).
In the case of the presence of hierarchical set of islands, \( \alpha  \) can
be defined through scaling properties of the island areas. In the considered
situation random walk is more or less uniform but trajectories are entangled
near stable/unstable manifolds, i.e. in the light gray areas of Figs. \ref{specialspeedsk02}-\ref{specialspeedk041}.
That means that \( \alpha \sim 2 \) although it is not exactly 2. Finally,
we arrive to: 
\begin{equation}
\label{equ5_{1}7}
\mu =2\beta /\alpha \sim 3/2
\end{equation}
 in correspondence to observations in the Table \ref{tablesec5}. The value
(\ref{equ5_{1}7}) was also discussed in \cite{ZasEdel2000} as one of possible
universal values for the transport exponent \( \mu  \).

\noindent We need to comment that it is not worthwhile to try to obtain \( \mu  \)
with a higher accuracy since a specific value of \( \mu  \) has no meaning
due to multi-fractal nature of transport \cite{ZasEdel2000}. It is also important
that we have considered such values of the control parameter \( k \) for which
there exists a strong filamentation. That guarantees a possibility of using
Eq. (\ref{equ5_{7}}) and the following analysis.

\section{Conclusion}

We have considered the dynamical and statistical properties of the passive particle
advection in a family of flows, created by three point vortices of different
signs. In all three particular cases, investigated numerically, tracer advection
was strongly chaotic: advection patterns, visualized via Poincare sections of
tracer trajectories are dominated by a well developed stochastic sea, occupying
most of the area around the center of vorticity. With the approach of the vortex
system to the collapse configuration, the degree of tracer chaotization increases:
the stochastic sea grows, expanding outward and consuming some of the inner
resonant islands.

The statistics of the tracers in the chaotic region is non-Gaussian. Anomalous
diffusion (faster than linear growth of variance) with different time and space
scales was found in all three cases, as well as non-Poissonian distributions
of Poincar\'{e} recurrences (with power law decay of long recurrence probability).
We did not find normal transport regimes, if such regime exist, they are confined
to narrow windows in the parameter domain.

Transport anomalies are caused by the phenomenon of stickiness of the chaotic
trajectories to the highly structured boundaries of the chaotic region. In the
cases considered, three important types of boundaries can be distinguished:
external border of the chaotic sea, boundaries of the resonant islands inside
the chaotic sea, and boundaries of the vortex cores. Each of these influences
various aspects of tracer statistics, analysis of their separate contributions
shows, that the vortex cores, that rotate with the fastest rate, determine the
high moments of the tracer distribution, while the external boundary, being
the slowest, but the most sticky, dominate the low moments.

Vortex cores appeared in simulations \cite{Babiano,NeufeldTel,KZ98.1}, their
origin and sizes were derived in \cite{KZ98.1} for a system of three identical
vortices; particularly it was shown that the cores are the islands of stability
filled by invariant curves and extremely thin stochastic layers. As the control
parameter \( k \) approaches the collapse value \( k_{c}=1/2 \), the sizes
of the vortex cores noticeably decrease. An upper bound of the core radii, obtained
from the minimum distance of vortex approach to each other, gives a good estimation
for both positive and negative vortex core size.

Although the transport possesses multi-fractal features, it can be successively
described by a fractional kinetic equation with characteristic exponents \( \alpha \sim 2 \)
and \( \beta \sim 3/2 \). A corresponding moments dependence is 
\begin{equation}
\label{equ6_{1}}
\langle |\theta |^{\alpha }\rangle \sim t^{\beta }\, \, .
\end{equation}
 The transport can be characterized by strong intermittency which manifests
itself in strong deviation from (\ref{equ6_{1}}) for higher moments, i.e. 
\begin{equation}
\label{equ6_{2}}
\langle |\theta |^{2m}\rangle \sim t^{\mu (m)}
\end{equation}
 with \( \mu \approx 3/2 \) for \( m=1 \) and \( \mu \approx 2m \) for large
values of \( m \). The latter corresponds to a strong influence of ballistic
regime of tracer dynamics.

We note that the value \( \beta \sim 3/2 \) for \( \alpha \sim 2 \) has also
been observed for flows generated by three identical vortices \cite{KZ2000}
and since this value remains for 3-vortex flows with extreme stress (vicinity
of collapse) we may reasonably speculate that for all periodic (bounded) three
vortex flows \( \beta \sim 3/2 \). We would like to point out that the present
work by analyzing the role played in transport by the different structures involved
in the flow using various techniques, and by confirming a typical value of the
second moment exponent should be of interest for the analysis of more realistic
and complicated systems involving many vortices and coherent structures such
as geophysical fluid dynamics.

\section*{\noindent Acknowledgments}

We would like to thanks N. J. Zabusky for usefull discussions regarding the
special influence of vortex collapse in passive tracers dynamics. 

This work was supported by the US Department of Navy, Grant No. N00014-96-1-0055,
and the US Department of Energy, Grant No. DE-FG02-92ER54184. This research
was supported in part by NSF cooperative agreement ACI-9619020 through computing
resources provided by the National Partnership for Advanced Computational Infrastructure
at the San Diego Supercomputer Center.

\appendix

\section*{Exponential Period Growth\label{appendix1}}

In this appendix, we recall some earlier results presented in \cite{LKZ2000},
and compute an asymptotic of the period growth as a function of \( \delta =1/2-k \).

It has been shown for the case of three vortices with two identical ones that
the relative motion of these vortices can been described using an one-dimensional
effective Hamiltonian 
\begin{equation}
\label{Hamilteff}
H_{eff}(\dot{X},X;\Lambda ,K,k)\equiv P^{2}/2+V(X)=0\: ,
\end{equation}
 with Hamiltonian equations 
\begin{equation}
\dot{X}=\partial H_{eff}/\partial P\equiv P\: ,\hspace {10mm}\dot{P}=-\partial H_{eff}/\partial X\: ,
\end{equation}
 where \( X=R^{2}_{1} \) is the square of the distance between the two positive
vortices, and the potential \( V \) has the following form 
\begin{equation}
\label{potential}
V(X)\equiv \frac{[(K-(1-k)X)^{2}-4k^{2}Y][(X-K)^{2}-4k^{2}Y]}{8\pi ^{2}k^{2}Y^{2}}\: ,\hspace {10mm}Y=(\Lambda X)^{1/k}\: .
\end{equation}
 Let us now estimate the period of the relative motion in our case. In this
paper we choose a situation with \( K=0 \), using then the transformation 
\begin{equation}
\label{defof U}
U=X^{2}/4k^{2}Y\: ,
\end{equation}
 the potential (\ref{potential}) becomes, 
\begin{equation}
\label{potforK=0}
V(U)=\frac{2}{\pi ^{2}}k^{2}\left( 1-k\right) ^{2}\left( U-\frac{1}{\left( 1-k\right) ^{2}}\right) \left( U-1\right) \: ,
\end{equation}
 given the fact that \( H_{eff}=0 \), the motion is confined to the negative
regions of the potential which leads to 
\begin{equation}
\label{rangefroK0}
1\leq U\leq \frac{1}{(1-k)^{2}}\: .
\end{equation}
 Consequently we obtain from (\ref{rangefroK0}) the boundaries for \( X \)
during the motion. We then compute the period of the relative motion using the
effective potential 
\begin{equation}
\label{perioddef}
T=2\int ^{X_{2}}_{X_{1}}\frac{dX}{\sqrt{-V(X)}}\: .
\end{equation}
 In the limit \( \delta \rightarrow 0 \) (\( k\rightarrow 1/2 \)), we obtain
\begin{equation}
\label{potubis}
V(U)\sim \frac{1}{8\pi ^{2}}(U-4)(U-1)\: ,
\end{equation}
 and using the inverse transformation, 
\begin{equation}
\label{reversedefofU}
X=(4k^{2}\Lambda ^{1/k}U)^{1/(2-1/k)}\sim (\Lambda ^{2}U)^{-1/4\delta }\: ,
\end{equation}
 we express the period in terms of \( U \), which leads to 
\begin{equation}
\label{periodu}
T\sim \frac{1}{4\delta }\int ^{4}_{1}\frac{(\Lambda ^{2}U)^{-1-1/4\delta }}{\sqrt{(U-1)(4-U)}}dU\: .
\end{equation}
 As \( \delta \rightarrow 0 \) (\( k\rightarrow 1/2 \)), it is the numerator
in (\ref{periodu}), which defines the asymptotic behavior, since it is decreasing
function of \( U \), the dominant term is from \( U=1 \) which leads to 
\begin{equation}
\label{periodfinalbound}
T\sim \frac{1}{\delta }\Lambda ^{-1/2\delta }\: .
\end{equation}
 Since the condition \( \Lambda =0.9<1 \), is verified (and is necessary for
the collapse to happen \cite{LKZ2000}), we have an exponential growth of the
period as the vortex-collapse configuration is approached.

\newpage

\newpage

\newpage

\newpage

\newpage

\newpage

\newpage

\newpage

\newpage

\newpage

\newpage

\newpage

\newpage

\newpage

\newpage

\newpage

\newpage

\newpage

\newpage

\newpage

\newpage 

\newpage

\newpage

\newpage

\begin{thebibliography}{10}
\bibitem{Aref84}H.Aref, \emph{Stirring by chaotic advection}, J. Fluid Mech. \textbf{143}, 1
(1984) 
\bibitem{Zaslav88}G. M. Zaslavsky, R. Z. Sagdeev, and A. A. Chernikov, Zhurn. Eksp. Teor. Fiz.
\textbf{94}, 102 (1988) (Soviet Physics, JETP \textbf{67}, 270 (1988) 
\bibitem{Ottino89}J. Ottino, \textit{The kinematics of Mixing: Stretching, Chaos, and Transport}
(Cambrige U. P., Cambrige, 1989) 
\bibitem{Aref90}H.Aref, \emph{Chaotic advection of fluid particles}, Phil. Trans. R. Soc. London
\textbf{A 333}, 273 (1990) 
\bibitem{Ottino90}J. Ottino, Mixing, \emph{Chaotic advection and turbulence}, Ann. Rev. Fluid
Mech. \textbf{22}, 207 (1990) 
\bibitem{RomKedar90}V. Rom-Kedar, A. Leonard and S. Wiggins, \emph{An analytical study of transport
mixing and chaos in an unsteady vortical flow}, J. Fluid Mech. \textbf{214},
347 (1990) 
\bibitem{Zaslav91}G. M. Zaslavsky, R. Z. Sagdeev, D. A. Usikov, and A. A. Chernikov, \emph{Weak
Chaos and Quasiregular Patterns}, Cambridge Univ. Press, (Cambridge 1991) 
\bibitem{Crisanti91}A. Crisanti, M. Falcioni, G. Paladin and A. Vulpiani, \emph{Lagrangian Chaos:
Transport, Mixing and Diffusion in Fluids}, La Rivista del Nuovo Cimento, \textbf{14},
1 (1991) 
\bibitem{Crisanti92}A. Crisanti, M. Falcioni, A. Provenzale, P. Tanga and A. Vulpiani, \emph{Dynamics
of passively advected impurities in simple two-dimensional flow models}, Phys.
Fluids A \textbf{4}, 1805 (1992) 
\bibitem{provenzale99}A. Provenzale, \emph{Transport by Coherent Barotropic Vortices}, Annu. Rev.
Fluid Mech. \textbf{31}, 55 (1999) 
\bibitem{Majda99}A. J. Majda, J. P. Kramer, \emph{Simplified models for turbulent diffusion:
Theory, numerical modeling, and physical phenomena}, Phys. Rep. \textbf{314},
238 (1999) 
\bibitem{Weiss98}J. B. Weiss, A. Provenzale, J. C. McWilliams, \emph{Lagrangian dynamics in high-dimensional
point-vortex system}, Phys. Fluids \textbf{10}, 1929 (1998)
\bibitem{mcwilliams99}J. C. McWilliams, J. B. Weiss and I Yavneh, \emph{The vortices of homogeneous
geostrophic turbulence}, J. Fuild Mech. \textbf{401}, 1 (1999)
\bibitem{Hardenberg2000}J. von Hardenberg, J. C. McWilliams, A. Provenzale, A. Shchepetkin and J. B.
Weiss, \emph{Vortex merging in quasi-geostrophic flows}, J. Fluid Mech. \textbf{412},
331 (2000)
\bibitem{Tabeling98}P. Tabeling, A.E. Hansen, J. Paret, \emph{Forced and Decaying 2D turbulence:
Experimental Study}, in \textit{``Chaos, Kinetics and Nonlinear Dynamics in
Fluids and Plasma'', eds. Sadruddin Benkadda and George Zaslavsky}, p. 145,
(Springer 1998) 
\bibitem{Hansen98}A. E Hansen, D. Marteau , P. Tabeling, \emph{Two-dimensional turbulence and
dispersion in a freely decaying system}, Phys. Rev. E \textbf{58}, 7261 (1998) 
\bibitem{Tabeling91}P. Tabeling, S. Burkhart, O. Cardoso, H. Willaime, \emph{Experimental-Study
of Freely Decaying 2-Dimensional Turbulence,} Phys. Rev. Lett. \textbf{67},
3772 (1991) 
\bibitem{Meleshko93}V.V. Meleshko, M.Yu. Konstantinov, \textit{Vortex Dynamics and Chaotic Phenomena},
(World Scientific, Singapore, 1999) 
\bibitem{Benzi86}R. Benzi, G. Paladin, S. Patarnello, P. Santangelo and A. Vulpiani, \textit{Intermittency
and coherent structures in two-dimensional turbulence}, J. Phys A \textbf{19},
3771 (1986) 
\bibitem{Benzi88}R. Benzi, S. Patarnello and P. Santangelo, \textit{Self-similar coherent structures
in two-dimensional decaying turbulence}, J. Phys A \textbf{21}, 1221 (1988) 
\bibitem{Weiss87}J. B. Weiss, J. C. McWilliams, \textit{Temporal scaling behavior of decaying
two-dimensional turbulence}, Phys. Fluids A \textbf{5}, 608 (1992) 
\bibitem{Bracco2000} A. Bracco, J. C. McWilliams, G. Murante, A. Provenzale, J. B. Weiss,
\textit{Revisiting freely decaying two-dimensional turbulence at millennial resolution}, Phys. Fluids 
\textbf{12} 2931, (2000)
\bibitem{McWilliams84}J. C. McWilliams, \textit{The emergence of isolated coherent vortices in turbulent
flow}, J. Fluid Mech. \textbf{146}, 21 (1984) 
\bibitem{McWilliams90}J. C. McWilliams, \textit{The vortices of two-dimensional turbulence}, J. Fluid
Mech. \textbf{219}, 361 (1990) 
\bibitem{Elhmaidi93}D. Elhma\"{\i}di, A. Provenzale and A. Babiano, \textit{Elementary topology
of two-dimensional turbulence from a Lagrangian viewpoint and single particle
dispersion}, J. Fluid Mech. \textbf{257}, 533 (1993) 
\bibitem{Carnevale91}G. F. Carnevale, J. C. McWilliams, Y. Pomeau, J. B. Weiss and W. R. Young, \textit{Evolution
of Vortex Statistics in Two-Dimensional Turbulence}, Phys. Rev. Lett. \textbf{66},
2735 (1991) 
\bibitem{Chernikov90}A. A. Chernikov, B. A. Petrovichev, A. V. Rogal'sky, R. Z. Sagdeev, and G. M.
Zaslavsky, \emph{Anomalous Transport of Streamlines Due to their Chaos and their
Spatial Topology}, Phys. Lett. A \textbf{144}, 127 (1990) 
\bibitem{Solomon94}T.H. Solomon, E.R. Weeks, H.L. Swinney, \emph{Chaotic advection in a two-dimensional
flow: L\'{e}vy flights in and anomalous diffusion}, Physica D, \textbf{76},
70 (1994) 
\bibitem{Weeks96}E.R. Weeks, J.S. Urbach, H.L. Swinney, \emph{Anomalous diffusion in asymmetric
random walks with a quasi-geostrophic flow example}, Physica D, \textbf{97},
219 (1996) 
\bibitem{Zaslavsky93}G.M. Zaslavsky, D. Stevens, H. Weitzner, \emph{Self-similar transport in incomplete
chaos}, Phys. Rev E \textbf{48}, 1683 (1993) 
\bibitem{Kovalyov2000}S. Kovalyov, \emph{Phase space structure and anomalous diffusion in a rotational
fluid experiment}, Chaos \textbf{10}, 153 (2000) 
\bibitem{zaslavsky93.2}G. M. Zaslavsky, \emph{How Long is the Way from Chaos to Turbulence?}, in ``\emph{New
Approaches and Concepts in Turbulence'', eds. Th. Dracos and A. Tsinober},
p 165 (Birkhäuser Verlag 1993)
\bibitem{Novikov78} E. A. Novikov, Yu. B. Sedov, Stochastic properties of a four-vortex system,
Sov. Phys. JETP \textbf{48}, 440 (1978)
\bibitem{ArefPomph80} H.Aref and N. Pomphrey, Integrable and chaotic motion of four vortices, Phys.
Lett. A \textbf{78}, 297 (1980)
\bibitem{Aref99}A. Aref, P. L. Boyland, M. A. Stremler, and D. L. Vainchtein \emph{Turbulent Statistical dynamics of a system of point vortices},
in \emph{``Trends in Mathematics''}, p 151 (Birkhäuser Verlag 1999)
\bibitem{Novikov75}E. A. Novikov, \emph{Dynamics and statistics of a system of vortices}, Sov.
Phys. JETP \textbf{41}, 937 (1975) 
\bibitem{Sire99}C. Sire and P. H. Chavanis, \emph{Numerical renormalization group of the vortex
aggregation in 2D decaying turbulence: the role of three-body interactions.}
cond-mat/9912222 (1999)
\bibitem{Zabusky96}D. G. Dritschel, N. J. Zabusky, \emph{On the nature of the vortex interactions
and models in unforced nearly-inviscid two-dimensional turbulence}, Phys. Fluids
\textbf{8}(5), 1252 (1996) 
\bibitem{KZ98.1}L.~Kuznetsov and G.M.~Zaslavsky, \emph{Regular and Chaotic advection in the
flow field of a three-vortex system}, Phys. Rev E \textbf{58}, 7330 (1998) 
\bibitem{KZ2000}L. Kuznetsov and G. M. Zaslavsky, \emph{Passive particle transport in three-vortex
flow}. To appear in Phys. Rev. E. 
\bibitem{LKZ2000}X. Leoncini, L. Kuznetsov and G. M. Zaslavsky, \emph{Motion of Three Vortices
near Collapse}, Preprint 1999 
\bibitem{Aref79}H. Aref, \textit{Motion of three vortices}, Phys. Fluids \textbf{22}, 393 (1979) 
\bibitem{NeufeldTel} Z. Neufeld and T. T\'{e}l, The vortex dynamics analogue of the restricted
three-body problem: advection in the field of three identical point vortices,
J. Phys. A: Math. Gen. \textbf{30}, 2263 (1997) 
\bibitem{Novikov79}E. A. Novikov, Yu. B. Sedov, \textit{Vortex collapse}, Sov. Phys. JETP \textbf{22},
297 (1979) 
\bibitem{Synge49}J. L. Synge, \textit{On the motion of three vortices}, Can. J. Math. \textbf{1},
257 (1949) 
\bibitem{Tavantzis88}J. Tavantzis and L. Ting, \textit{The dynamics of three vortices revisited},
Phys. Fluids \textbf{31}, 1392 (1988) 
\bibitem{Kimura90}Y. Kimura, \textit{Parametric motion of complex-time singularity toward real
collapse}, Physica D \textbf{46}, 439 (1990) 
\bibitem{Machioro94}C. Machioro and M. Pulvirenti, \textit{Mathematical theory of incompressible
nonviscous fluids}, Applied Mathematical Science 96 (Springer-Verlag, New York,
1994) 
\bibitem{Saffman95}P. Saffman, \textit{Vortex Dynamics}, Cambridge Monographs on Mechanics and
Applied Mathematics (Cambridge University Press, Cambridge, 1995) 
\bibitem{McLachlan92}R.I. McLachlan, P. Atela, \emph{The accuracy of symplectic integrators}, Nonlinearity
\textbf{5}, 541 (1992) 
\bibitem{Zaslavsky92}G.M. Zaslavsky, in \textit{``Topological Aspects of the Dynamics of Fluids
and Plasmas'', ed. H.K. Moffatt, et al}, p. 481, (Kluwer, Dordrecht, 1992);
Chaos \textbf{4}, 25 (1994); Physica D \textbf{76}, 110 (1994) 
\bibitem{Chirikov79}B. V. Chirikov, Phys. Rep. \textbf{52}, 264 (1979) 
\bibitem{Rechester80}A.B. Rechester, R. White, Phys. Rev. Lett. \textbf{44}, 1586 (1980) 
\bibitem{dCN98}D. del-Castillo-Negrete, \emph{Asymmetric transport and non-Gaussian statistics
of passive scalars in vortices in shear}, Phys. Fluids \textbf{10}, 576 (1998) 
\bibitem{Castiglione99}P.~Castiglione, A.~Mazzino, P.~Muratore-Ginanneschi and A.~Vulpiani, \emph{On}
\textit{strong} \emph{anomalous diffusion}, Physica D \textbf{134}, 75 (1999) 
\bibitem{Benkadda99}S. Benkadda, S. Kassibrakis, R.B. White, G.M. Zaslavsky, Phys. Rev. E, \textbf{59},
3761 (1999) 
\bibitem{zasrep}G. M. Zaslavsky, and B. A. Niyazov, Phys. Rep. \textbf{283}, 73 (1997) 
\bibitem{ZEN97}G. M. Zaslavsky, M. Edelman, B. A. Niyazov, \emph{Self-similarity, renormalization,
and phase space nonuniformity of Hamiltonian chaotic dynamics}, Chaos \textbf{7},
159 (1997) 
\bibitem{ZasEdel2000}G. M. Zaslavsky and M. Edelman, \emph{Hierachical structures in the phase space
and fractional kinetics: I classical systems}, Chaos \textbf{10}, 135 (2000) 
\bibitem{Hentschel84}H. G. E. Hentschel and I. Procaccia, Physica D \textbf{8}, 435 (1983); P. Grassberger
and I. Procaccia, \emph{ibid}, \textbf{13}, 34 (1984) 
\bibitem{Frish85}U. Frisch and G. Parisi, in \emph{Turbulence and Predictability of Geophysical
Flows and Climate Dynamics}, edited by M. Ghill, R. Benzi, and G. Parisi (North-Holland,
Amsterdam, 1985) 
\bibitem{Jensen85}M. H. Jensen, L. P. Kadanoff, A. Libshaber, I. Procaccia, and J. Stavans, Phys.
Rev. Lett. \textbf{55}, 439 (1985); T. C. Halsey, M. H. Jensen, L. P. Kadanoff,
A. Libshaber, I. Procaccia, and B. I. Schraiman, Phys. Rev. A \textbf{33}, 1141
(1986) 
\bibitem{Melnikov96}V. K. Melnikov, \emph{On the existence of self-similar structures in the resonnance
domain}, in \emph{Transport, Chaos and Plasma Physics} \emph{II}, \emph{Proceedings,
Marseille}, edited by F. Doveil, S. Benkadda, and Y. Elskens (World Scientific,
Singapore, 1996), pp. 142-153 
\bibitem{Karney83}C. C. F. Karney, Physica D \textbf{8}, 360 (1983) 
\bibitem{Rom-Kedar99}V. Rom-Kedar and G. M. Zaslavsky, \emph{Islands of accelerator modes and homoclinic
tangles}, Chaos \textbf{9}, 697 (1999) 
\bibitem{Montroll84}E. W. Montroll and M. F. Shlesinger, in \emph{Studies in Statistical Mechanics},
edited by J. Lebowitz and E. Montroll (North-Holland, Amsterdam, 1984), Vol.
11, p. 1 
\bibitem{Lamb45}H. Lamb, \textit{Hydrodynamics}, (6th ed. New York, Dover, 1945) 
\bibitem{Babiano}A. Babiano, G. Boffetta, A. Provenzale and A. Vulpiani, \emph{Chaotic advection
in point vortex models and two-dimensional turbulence}, Phys. Fluids \textbf{6},
2465 (1994)
\end{thebibliography}
\end{document}